\documentclass[aps,prd,nofootinbib,floatfix,superscriptaddress,secnumarabic,preprintnumbers,longbibliography]{revtex4-2}
\usepackage{amsmath}
\usepackage{amsfonts}
\usepackage{graphicx}
\usepackage{subcaption}
\usepackage{xcolor}
\usepackage{commath}
\usepackage{hyperref}
\usepackage[normalem]{ulem}

\def\slashed{{/}\mskip-10.0mu}
\def\pcircslash{\slashed {p\mskip -5mu ^{^\circ}}}
\def\pcirc{ {p\mskip -5mu ^{^\circ}}}
\def\archcoth{\rm arccoth}

\begin{document}

\title{Perturbative determination of $\mathcal{O}(am)$ improvement on the QCD running coupling}

\author{M.~Costa}
\email[]{marios.costa@cut.ac.cy}
\affiliation{Department of Physics, University of Cyprus, Nicosia, CY-1678, Cyprus}
\affiliation{Present address: Department of Mechanical Engineering and Material Science and Engineering, Cyprus University of Technology, Limassol, CY-3036, Cyprus}
\affiliation{Rinnoco Ltd, Limassol, CY-3047, Cyprus}

\author{D.~Gavriel}
\email[]{demetrianos.gavriel@unipr.it}
\affiliation{Department of Physics, University of Cyprus, Nicosia, CY-1678, Cyprus}
\affiliation{Present address: Dipartimento di Scienze Matematiche, Fisiche e Informatiche, Universit\`a di Parma and INFN, Gruppo Collegato di Parma, I-43100 Parma, Italy}

\author{H.~Panagopoulos}
\email[]{panagopoulos.haris@ucy.ac.cy}
\affiliation{Department of Physics, University of Cyprus, Nicosia, CY-1678, Cyprus}

\author{G.~Spanoudes}
\email[]{spanoudes.gregoris@ucy.ac.cy}
\affiliation{Department of Physics, University of Cyprus, Nicosia, CY-1678, Cyprus}

\begin{abstract}
    We present the perturbative results of the discretization errors proportional to the quark mass [$\mathcal{O}(a m)$] on the QCD running coupling within lattice perturbation theory. Our analysis involves calculating the two-loop renormalization factor $Z_g$ using improved lattice actions for the $SU(N_c)$ gauge group and $N_f$ multiplets of fermions with a finite quark mass. We employ the background field method to compute $Z_g$, by calculating quantum corrections on both the background and quantum gluon propagator, respecting the $\mathcal{O}(a)$ improvement. This allows us to evaluate the perturbative $\mathcal{O}(a m)$ lattice errors which affect the determination of the running coupling. Eliminating these $\mathcal{O}(a m)$ effects is crucial for the nonperturbative studies of precision determinations of the strong coupling constant using lattice field theory.
\end{abstract}

\maketitle

\section{Introduction}

The world average of the strong-coupling constant ($\alpha_s$), evaluated at the Z-boson mass scale, includes various determinations, each classified based on their methodological approach (e.g, hadron collisions, $\tau$ decays, etc.). The most accurate determinations arise from lattice QCD, providing a result of $\alpha_s(m_Z^2) = 0.1183(7)$ \cite{FlavourLatticeAveragingGroupFLAG:2024oxs}. However, a number of systematic uncertainties arise in lattice QCD, associated with perturbative truncation, discretization errors in continuum-limit extrapolations, and finite-volume effects; all of these uncertainties have been taken into account in the above result.

Overcoming the limitations of simulations is challenging without employing a specific strategy to control systematic uncertainties. By applying the decoupling method~\cite{DallaBrida:2019mqg} with the finite-size-scaling technique, one can bridge the gap from low to high energies in a controlled manner. The decoupling method shifts the bulk of the calculation to pure gauge theory, where it is computationally cheaper. This combined approach effectively eliminates systematic errors in a controlled way, as opposed to most “large volume” approaches.

The decoupling strategy relies, to some extent, on lattice QCD computations with $N_f \geq 3$ degenerate heavy quarks at a low energy scale requiring the introduction of massive finite volume couplings. When linking observables at different quark masses, it is important to keep a constant lattice spacing up to order $\mathcal{O}(a^2)$, and hence there is a need to study a lattice theory that respects $\mathcal{O}(a)$ improvements.

In an $\mathcal{O}(a)$ improved lattice theory, one has to properly impose renormalization conditions in a way that correlation functions of the renormalized fields converge to the continuum limit as $\mathcal{O}(a^2)$. A simple strategy is to choose all renormalization conditions defined at the same point ($g_0$, $a m_0$) in the bare parameter space. In this scheme, transformations of the bare parameters and rescaling of the bare fields do not affect renormalization quantities. Hence, the corresponding counterterms of $\mathcal{O}(a)$ can be ignored and so this category of renormalization schemes is automatically compatible with $\mathcal{O}(a)$ improvement. However, in this manner, the renormalized coupling constant and renormalized fields implicitly depend on the quark mass. This can be avoided though, by using mass-independent renormalization schemes \cite{Luscher1996}.

In mass-independent schemes, renormalization conditions are defined at zero quark mass. As a result, renormalized quark masses are decoupled from the running coupling and it is convenient to study its scale evolution ($\beta$-function) since the arguments of the massless theory will remain the same. Then, the issue is that the theory of finite mass quarks must be related to the massless theory. This link is usually established through the bare parameters. Thus, there is a need for reparametrization of the bare theory, so that we can preserve the $\mathcal{O}(a)$ improvement.

To set up a general mass-independent renormalization scheme respecting $\mathcal{O}(a)$ improvement, one can introduce a modified bare coupling through
\begin{equation}
    \Tilde{g}_0^2= g_0^2 (1 + b_g(g_0^2) \:a m_q)
    \label{eq:modified_g0}
\end{equation}
while the subtracted degenerate quark mass, $m_q$, is given by
\begin{equation}
    m_q=m_0-m_c(g_0^2)
    \label{eq:subtracted mass}
\end{equation}

The parameter $m_c(g_0^2)$ is the critical value of the Lagrangian quark mass at which the renormalized quark masses vanish. Note that $m_c(g_0^2)$ depends on how the physical quark mass is defined, but its exact form should differ only at $\mathcal{O}(a^2)$. Nevertheless, order $a^2$ corrections are considered negligible. At $m_0 = m_c$, the modified and ordinary bare couplings coincide. Moreover, $b_g(g^2_0)$ in Eq.~(\ref{eq:modified_g0}) must be appropriately selected so as to remove any remaining cutoff effects of $\mathcal{O}(a)$. The bottom line is that the scaling required for $g_0$ depends on the quark mass, while $\Tilde{g}_0$ scales independently of the quark mass.

For heavier quarks in decoupling methods, values of $a m_q$ increase, emphasizing the significance of determining $b_g(g_0^2)$ in both perturbative and nonperturbative ways. Nonperturbative calculations of $b_g(g_0^2)$ have been recently carried out \cite{DallaBrida:2023fpl}. Currently, $b_g(g_0^2)$ is only known to one-loop order in perturbation theory \cite{Luscher1996}, which introduces a significant systematic error to the precision of the strong coupling constant due to the truncated perturbative result \cite{DallaBrida:2019mqg}.

In the present work, we study the two-loop renormalization factor, $Z_g$ . This factor relates the bare running coupling $g_0$ to the $\rm \overline{MS}$-renormalized running coupling $g$. By employing the background field method we find the discretization errors associated with the finite quark mass. We are specifically interested in identifying discretization errors that are proportional to the quark mass. These errors are of the order of $\mathcal{O}(a m)$ and can affect the determination of the strong coupling constant. To calculate $b_g(g_0^2)$ in Eq.~(\ref{eq:modified_g0}) at two-loop order, which has not been done before, we need to take into account the complexity caused by the inclusion of quark masses \cite{Dietrich:2009ns}. Eliminating $\mathcal{O}(a m)$ effects is important for improving the accuracy of any quantity that is calculated using Wilson-type fermions \cite{DallaBrida:2022eua}.

The paper is organized as follows: In Secs.~\ref{sec:lattice_actions} and~\ref{sec:theoretical_setup}, we provide the setup of our calculation, including the improved lattice actions, and a brief review of the beta function and the background field formalism. Sections~\ref{sec:1loop} and~\ref{sec:2loop} present our main results at the one- and two-loop order, respectively. This includes expressions for both the Green's functions and $b_g(g_0^2)$. In Sec.~\ref{sec:conclusions}, we summarize our findings and outline future plans. Appendix~\ref{ap:integration} briefly describes integration over loop momenta and estimation of the corresponding numerical error; Appendix~\ref{ap:tables} contains tables with the numerical values of the two-loop lattice Green's function of each Feynman diagram. 

\section{Improved lattice action}
\label{sec:lattice_actions}

Our computations are carried out within the lattice regularization, using the clover improved action (Sheikholeslami-Wohlert) \cite{Sheikholeslami1985} for fermions. The clover action reads, in standard notation,
\begin{eqnarray}
S_{\rm F} &=& \sum_{f}\sum_{x} (4r+m_0)\bar{\psi}_{f}(x)\psi_f(x)\nonumber \\
&-& \frac{1}{2}\sum_{f}\sum_{x,\,\mu}\bigg{[}\bar{\psi}_{f}(x) \left( r - \gamma_\mu\right)
U_{x,\, x+\mu}\psi_f(x+{\mu}) 
+\bar{\psi}_f(x+{\mu})\left( r + \gamma_\mu\right)U_{x+\mu,\,x}\psi_{f}(x)\bigg{]}\nonumber \\
&-& \frac{1}{4}\,c_{\rm sw}\,\sum_{f}\sum_{x,\,\mu,\,\nu} \bar{\psi}_{f}(x)
\sigma_{\mu\nu} {\hat F}_{\mu\nu}(x) \psi_f(x),
\label{eq:fermion_action}
\end{eqnarray}
where $r$ is the Wilson parameter, $f$ is a flavor
index, $\sigma_{\mu \nu} = [\gamma_{\mu},\gamma_{\nu}]/2$, $m_0$ is the Lagrangian quark mass,\footnote{For simplicity of notation, we denote all flavor masses by $m_0$; the case of different flavor masses can be trivially recovered from our results.} and $c_{\rm sw}$ is the clover parameter.  In the following calculations, $r$ is set to 1 as customary, and $c_{\rm sw}$ is considered as a free parameter for wider applicability of results. Powers of the lattice spacing $a$ have been omitted and may be directly reinserted by dimensional counting. The tensor $\hat{F}_{\mu\nu}$ is proportional to a lattice representation of the gluon field tensor, defined through
\begin{equation}
    {\hat F}_{\mu\nu} \equiv \frac{1}{8}\,(Q_{\mu\nu} - Q_{\nu\mu})
\end{equation}
where $Q_{\mu\nu}$ is the sum of the plaquette loops:
\begin{eqnarray}
Q_{\mu\nu} &=& U_{x,\, x+\mu}U_{x+\mu,\, x+\mu+\nu}U_{x+\mu+\nu,\, x+\nu}U_{x+\nu,\, x} \nonumber \\
&+& U_{ x,\, x+ \nu}U_{ x+ \nu,\, x+ \nu- \mu}U_{ x+ \nu- \mu,\, x- \mu}U_{ x- \mu,\, x} \nonumber \\
&+& U_{ x,\, x- \mu}U_{ x- \mu,\, x- \mu- \nu}U_{ x- \mu- \nu,\, x- \nu}U_{ x- \nu,\, x} \nonumber \\
&+& U_{ x,\, x- \nu}U_{ x- \nu,\, x- \nu+ \mu}U_{ x- \nu+ \mu,\, x+ \mu}U_{ x+ \mu,\, x}
\end{eqnarray}

For the gauge fields we employ a class of Symanzik improved gauge actions \cite{Horsley:2004mx}, involving Wilson loops with 4 and 6 links ($1\times1$ plaquettes and $1\times2$ rectangles, respectively), which is given by the relation
\begin{equation}
S_{\rm G}=\frac{2}{g_0^2} \left[ c_0 \sum_{\rm plaq.} {\rm Re\,Tr\,}\{1-U_{\rm plaq.}\} 
+ c_1 \sum_{\rm rect.} {\rm Re \, Tr\,}\{1- U_{\rm rect.}\} \right]
\label{eq:gluon_action}
\end{equation}
The coefficients $c_0$ and $c_1$ can in principle be chosen arbitrarily, subject to the following normalization condition, which ensures the correct classical continuum limit of the action
\begin{equation}
c_0 + 8 c_1  = 1.
\label{norm}
\end{equation}
For the numerical evaluation, particular choices of values for $\{c_0$, $c_1\}$ are employed in our calculations as shown in Table \ref{tb:Symanzik_coeff}.

\begin{table}[ht]
\begin{center}
\begin{tabular}{|l|cc|}
\hline
Gluon action   &  $c_0$	  & $c_1$ \\
\hline
Wilson         &	$1$       &	$0$       \\
Tree-level Symanzik (TLS)$\;\;$    &	$5/3$     & $-1/12$   \\
Iwasaki	       &   $\;\;3.648\;\;$    & $\;\;-0.331\;\;$  \\
\hline
\end{tabular}
\end{center}
\caption{Commonly used sets of values for Symanzik coefficients.}
\label{tb:Symanzik_coeff}
\end{table}

\section{Theoretical setup}
\label{sec:theoretical_setup}

Let us first recall some concepts related to the beta function in the massless case and to the background field method; these will be useful as we extend them to the case of nonzero fermion mass. The dependence of the renormalized coupling constant on the intrinsic scale of the renormalization scheme is given by the renormalized $\beta$-function:
\begin{equation}
     \beta(g) \equiv \bar{\mu} \frac{dg}{d\bar{\mu}} \Big|_{a,g_0}
\label{eq:beta_function_DR}
\end{equation}
where $a$ is the lattice spacing, $\bar{\mu}$ is the renormalization scale and $g$ ($g_0$) is the renormalized (bare) coupling constant. We will employ the $\rm \overline{MS}$ renormalization scheme in this work. In the asymptotic limit for QCD ($g \rightarrow 0$), one can write the expansion of the $\beta$-function in powers of $g$:
\begin{equation}
    \beta(g) =\,  -b_0 \,g^3 -b_1 \,g^5 - b_2\,g^7 + \mathcal{O} (g^9)
    \label{eq:beta_function_DR_expansion}
\end{equation}

The bare $\beta$-function for the lattice regularization is defined as
\begin{equation}
    \beta_L(g_0) \equiv -a \frac{dg_0}{da} \Big|_{\bar{\mu}, g } 
\label{eq:beta_function_LR}
\end{equation}
Similar to Eq.~(\ref{eq:beta_function_DR_expansion}), the asymptotic high energy limit ($g_0 \rightarrow 0$) of the lattice bare $\beta$-function is
\begin{equation} 
    \beta_L (g_0) = -b_0 {g_0}^3 - b_1 {g_0}^5 - b_2^L {g_0}^7 + \mathcal{O} ({g_0}^9)
\label{eq:beta_function_LR_expansion}
\end{equation}

The first two coefficients $b_0$, $b_1$ are universal for massless gauge theories (\cite{Balian1981}), in the sense that they do not depend on the regularization nor the renormalization scheme. However, coefficients $b_i^L (i \geq 2)$ depend on the renormalization scheme and must be determined perturbatively. Coefficients $b_0$, $b_1$, and $b_2$ can be found in Ref.~\cite{Tarasov1980}, while $b_2^L$ has been calculated either in the absence of fermions \cite{Luscher1995,Alles:1996cy} or using the Wilson \cite{Christou:1998ws}, clover \cite{Bode:2001uz}, or overlap \cite{Constantinou:2007rm} fermions.

The bare coupling constant $g_0$ is related to the renormalized coupling constant $g$ through the renormalization function $Z_g$ :
\begin{equation}
    g_0 = Z_g (g_0^2, a \bar{\mu}) g
    \label{eq:Zg}
\end{equation}

The most convenient and economical way to proceed with the calculation of $Z_g(g_0^2,a\bar{\mu})$ is to use the background field (BF) technique~\cite{Ellis1984, Luscher1995background}, in which the renormalization of the background field is directly connected with $Z_g$. On the lattice, the background field technique can be approached in more than one way. Different lattice actions may be chosen and the precise way in which the background field is introduced is arbitrary to some extent. However, the differences between the choices of lattice actions should be irrelevant in the continuum limit. The background field on the lattice is introduced by decomposing the gauge link variable as follows \cite{Ellis1984}:
\begin{equation} 
\begin{split}
    U_\mu (x) &= U^Q_{\mu} (x) U^B_{\mu} (x),  \\
    U^Q_{\mu} (x) &\equiv e^{i g_0 Q_{\mu} (x)},  \\
    U^B_{\mu} (x) &\equiv e^{i a B_{\mu} (x)}
\end{split}
\end{equation} 
where $Q_\mu (x) = Q_\mu^a (x) T^a$ is the quantum field, $B_\mu (x) = B_\mu^a (x) T^a$ is the background field, and $T^a$ are the generators of $SU(N_c)$ with the convention of $ {\rm Tr}(T^a T^b) = \delta^{ab}/2$. Since the gauge link is now a product of two different field links, there is freedom in interpreting the gauge transformation:
\begin{equation}
    {[U_\mu (x)]}^{\Lambda} = \Lambda (x) U_\mu (x) \Lambda^{-1} (x + a \hat{\mu}).
\end{equation}

 This transformation can be viewed in two ways. The first one considers the quantum field as a matter field which transforms purely locally, while the background field transforms as a true gauge field:
\begin{equation}
\begin{split}
    {[U^Q_\mu (x)]}^{\Lambda} &= \Lambda (x) U^Q_\mu (x) \Lambda^{-1} (x)  \\
    {[U^B_\mu (x)]}^{\Lambda} &= \Lambda (x) U^B_\mu (x) \Lambda^{-1} (x + a \hat{\mu})
\end{split}
\end{equation}

The second one considers the background field as invariant, while the quantum field is now the true gauge field:
\begin{equation}
\begin{split}
    {[U^Q_\mu (x)]}^{\Lambda} &= \Lambda (x) U^Q_\mu (x) U^B_\mu (x) \Lambda^{-1} (x + a \hat{\mu}) (U^B_\mu (x))^{-1} \\
    {[U^B_\mu (x)]}^{\Lambda} &= U^B_\mu (x)
\end{split}
\end{equation}
Let us call the first interpretation of gauge transformations ``background gauge transformations'' and the second one ``quantum gauge transformations.'' As the background is an external field, which is not involved in the path integration, the gauge-fixing term, which ensures the finiteness of path integrals, can be chosen to preserve the gauge invariance under background transformations. A proper choice is the following:
\begin{equation}
    S_{gf} = \lambda_0 a^4 \sum_{x, \mu , \nu} {\rm Tr} \, \left\{ D^-_{\mu} Q_{\mu}(x) D^-_{\nu} Q_{\nu}(x) \right\}
\label{eq:gauge_fixing_action}
\end{equation}
where $\lambda_0$ is the inverse bare gauge parameter [$\lambda_0^{-1} = 0 ~(1)$ in Landau (Feynman) gauge)] and the lattice covariant derivative $D^-_{\mu}$ is written as
\begin{equation}
    D^-_{\mu} Q_{\nu}(x) = \frac{1}{a} \Big[{U^{B}_{\mu}}^{-1}(x - a {\hat \mu}) \ Q_{\nu}(x - a {\hat \mu}) \ U^{B}_{\mu}(x - a {\hat \mu}) - Q_{\nu}(x) \Big]
\end{equation}
This term preserves the gauge invariance of the background field in the action, even though it breaks the gauge invariance of the quantum field.

The corresponding Faddeev-Popov action ($S_{\rm FP}$) for ghost fields, produced by the gauge-fixing term, along with the contribution from the integration measure ($S_{ \rm meas.}$), resulting from the change of integration variables from links to vector fields, are explicitly written in Ref.~\cite{Ellis1984}. Therefore the full action is given by
\begin{equation}
    S = S_{\rm F} + S_{\rm G} + S_{\rm gf} + S_{\rm FP} + S_{\rm meas.}
    \label{eq:full_action}
\end{equation}

The fact that exact gauge invariance is preserved in the background field formalism, leads to a relation between the renormalization factors of background field $Z_B$ and of coupling constant $Z_g$ \cite{Abbott1980}:
\begin{equation}
    Z_B (g_0^2, a \mu) \ Z_g^2 (g_0^2, a \mu)= 1
    \label{eq:ZB_Zg}
\end{equation} 
where $B^{\mu}_0 =  Z_B (g_0^2, a \mu)^{1/2} B^{\mu}$. In this framework, the relation between the lattice bare coupling constant and the renormalized one can be extracted by the evaluation of $Z_B$, instead of $Z_g^2$, with no need to calculate any 3-point Green's functions, thus leading to a less complicated calculation. Note,  however, that the inclusion of quark masses adds a layer of complexity.

\medskip
Henceforth we consider the 1-particle-irreducible (1-PI) 2-point Green's function of background field, both in the continuum [$\Gamma^{\rm BB}_{\rm R} (p,m,\lambda)_{\mu \nu}^{a b}$] and on the lattice [$\Gamma^{\rm BB}_{\rm L} (p,m_q,\lambda_0)_{\mu \nu}^{a b}$], in the presence of a fermion mass. These functions can be expressed in terms of scalar amplitudes $\nu_R (p,m,\lambda)$, $\nu (p,m_q,\lambda_0)$. Following the notation of Ref.~\cite{Luscher1995} the Green's functions of the background field in the continuum are given as
\begin{equation} 
\begin{split}
    \Gamma^{\rm BB}_{\rm R} (p,m,\lambda)_{\mu \nu}^{a b} &= - \delta^{a b} ( \delta_{\mu \nu} p^2 - p_{\mu} p_{\nu}) \left(1 - \nu_{\rm R} (p,m, \lambda)\right) / g^2  \;,  \\
    \nu_{\rm R} (p,m,\lambda) &= \sum_{\ell = 1}^\infty g^{2 \ell} \nu_{\rm R}^{(\ell)} (p,m(g_0^2),\lambda(g_0^2))
    \label{eq:gamma_BB_R}
\end{split} 
\end{equation} 
where $\nu_{\rm R}^{(\ell)} (p,m(g_0^2),\lambda(g_0^2))$ can be written as
\begin{equation}
    \nu_{\rm R}^{(\ell)} (p,m(g_0^2),\lambda(g_0^2)) = \nu_{\rm R}^{(\ell)} (p,0,\lambda(g_0^2)) + \Delta\nu_{\rm R}^{(\ell)} (p,m(g_0^2),\lambda(g_0^2))
    \label{eq:nu_R_ell} 
\end{equation}
where $\lambda$ is the inverse $\rm \overline{MS}$-renormalized gauge parameter, $m$ is the renormalized mass [where $m = Z_{m}(g_0^2) \ m_q$] and $\nu_{\rm R}^{(\ell)} (p,0,\lambda(g_0^2))$ is the amplitude corresponding to the massless case.

Similarly, the Green's functions of the background field on the lattice are given as
\begin{equation} 
\begin{split}
    \sum_\mu \Gamma^{\rm BB}_{\rm L} (p,m_q, \lambda_0)_{\mu \mu}^{a b} &= - \delta^{a b} 3 \hat{p}^2 \left(1 - \nu (p,m_q, \lambda_0)\right) / g^2_0  \;,  \\
    \nu(p,m_q,\lambda_0) &= \sum_{\ell = 1}^\infty g_0^{2 \ell} \nu^{(\ell)} (p,m_q,\lambda_0)
    \label{eq:gamma_BB_L}
\end{split} 
\end{equation} 
where $\hat{p}^2 = \sum_\mu \hat{p}^2_\mu$, $\hat{p}_\mu = (2 / a) \sin(a p_\mu / 2)$. The amplitude $\nu^{(\ell)} (p,m_q,\lambda_0)$ can be written as
\begin{align}
\begin{split}
    \nu^{(\ell)} (p,m_q,\lambda_0) &= \nu^{(\ell)}_0 (p,m_q,\lambda_0) +   a\,m_q \; \nu^{(\ell)}_1 (p,m_q,\lambda_0)  + {\cal O}(a^2\,m_q^2) \\
    &=\nu^{(\ell)}_0 (p,0,\lambda_0)+ \Delta\nu^{(\ell)}_0 (p,m_q,\lambda_0) +   a\,m_q \left( \nu^{(\ell)}_1 (p,0,\lambda_0) + \Delta\nu^{(\ell)}_1 (p,m_q,\lambda_0) \right) + {\cal O}(a^2\,m_q^2) 
    \label{eq:nu_L_ell} 
\end{split}
\end{align}
where $\nu^{(\ell)}_0 (p,0,\lambda_0)$ represents the outcome to the massless theory without considering the $\mathcal{O}(a)$ improvement. The tensor structure of these Green's functions, as given above, is implied by the symmetries of the theory.

As discussed in Ref.~\cite{Luscher1995background}, the continuum and lattice Green's functions are related by $\Gamma^{\rm BB}_{\rm R} =\Gamma^{\rm BB}_{\rm L}+ \mathcal{O} (a)$. Thus, using Eqs.~(\ref{eq:ZB_Zg}),(\ref{eq:gamma_BB_R}) and (\ref{eq:gamma_BB_L}), we can express $Z_g^2$ in terms of $\nu_R (p)$, $\nu (p)$:
\begin{equation}
    Z_g^2 = Z_B^{-1} = \frac{1 - \nu (p,m_q,\lambda_0)}{1 - \nu_{\rm R} (p,m,\lambda)} 
    \label{eq:Zg_nu} 
\end{equation}

Given that $Z_g^{\overline{\rm MS}}$ is gauge independent, we choose to compute it in the renormalized Feynman gauge, that is, we set the renormalized gauge parameter $\lambda = 1$. At the two-loop level, this necessitates an explicit renormalization of $\lambda_0$ to one loop. A similar relation to Eq.~(\ref{eq:Zg_nu}) for the renormalization factor $Z_{\lambda}$ [where $\lambda = Z_{\lambda}(g_0^2) \ \lambda_0$], can be expressed in terms of the scalar terms $\omega_{\rm R} (p,m)$ and $\omega (p,m_q)$ which appear in the definition of the quantum field self-energy in the continuum [$\Gamma^{\rm QQ}_{\rm R} (p,m,\lambda)_{\mu \nu}^{a b}$] and on the lattice [$\Gamma^{\rm QQ}_{\rm L} (p,m_q,\lambda_0)_{\mu \nu}^{a b}$], respectively: 
 
 \begin{equation} 
 \begin{split}
    \Gamma^{\rm QQ}_{\rm R} (p,m,\lambda)_{\mu \nu}^{a b} &= - \delta^{a b} \left[( \delta_{\mu \nu} p^2 - p_{\mu} p_{\nu}) \left(1 - \omega_{\rm R} (p,m)\right) + \lambda \ p_\mu p_\nu \right] \;,  \\
    \omega_{\rm R} (p,m) &= \sum_{\ell = 1}^\infty g^{2 \ell} \omega_{\rm R}^{(\ell)} (p,m(g_0^2))
    \label{eq:gamma_QQ_R}
\end{split} 
\end{equation} 
where $\omega_{\rm R}^{(\ell)} (p,m(g_0^2))$ can be written as
\begin{equation}
    \omega_{\rm R}^{(\ell)} (p,m(g_0^2)) = \omega_{\rm R}^{(\ell)} (p,0) + \Delta\omega_{\rm R}^{(\ell)} (p,m(g_0^2))
    \label{eq:omega_R_ell} 
\end{equation}
and $\omega_{\rm R}^{(\ell)} (p,0)$ is the amplitude corresponding to the massless case.

The Green's functions of the quantum field in the lattice are given as
\begin{equation} 
\begin{split}
    \sum_\mu \Gamma^{\rm QQ}_{\rm L} (p,m_q,\lambda_0)_{\mu \mu}^{a b} &= - \delta^{a b} \hat{p}^2 \left[3 \left(1 - \omega (p,m_q)\right) + \lambda_0\right]  \;,  \\
    \omega (p,m_q) &= \sum_{\ell = 1}^\infty g_0^{2 \ell} \omega^{(\ell)} (p,m_q)
    \label{eq:gamma_QQ_L}
\end{split} \end{equation} 
where the amplitude $\omega^{(\ell)} (p,m_q)$ can be written as
\begin{align}
\begin{split}
    \omega^{(\ell)} (p,m_q) &= \omega^{(\ell)}_0 (p,m_q) +   a\,m_q \; \omega^{(\ell)}_1 (p,m_q) + {\cal O}(a^2\,m_q^2) \\
    &=\omega^{(\ell)}_0 (p,0) + \Delta\omega^{(\ell)}_0 (p,m_q) +   a\,m_q \left(  \omega^{(\ell)}_1 (p,0) + \Delta\omega^{(\ell)}_1 (p,m_q) \right) + {\cal O}(a^2\,m_q^2) 
    \label{eq:omega_L_ell} 
\end{split}
\end{align}
and $\omega_0^{(\ell)} (p,0)$ is the amplitude corresponding to the massless case.

The relation between the continuum and lattice Green's functions is now given by $\Gamma^{\rm QQ}_{\rm R} = Z_{\lambda}^{-1} \Gamma^{\rm QQ}_{\rm L}+ \mathcal{O} (a)$ \cite{Luscher1995background}. Hence, using Eqs.~(\ref{eq:gamma_QQ_R}), and (\ref{eq:gamma_QQ_L}) we can express $Z_{\lambda}$ in terms of $\omega_R (p,m)$, $\omega (p,m_q)$:
\begin{equation}
    Z_{\lambda} = \frac{1 - \omega (p,m)}{1 - \omega_R (p,m_q)} 
    \label{eq:Zlambda_omega} 
\end{equation}

Expressed in terms of the perturbative expansions given by Eqs.~(\ref{eq:gamma_BB_R}), (\ref{eq:gamma_BB_L}), (\ref{eq:gamma_QQ_R}), and (\ref{eq:gamma_QQ_L}), and utilizing the relations provided by Eqs.~(\ref{eq:Zg_nu}) and (\ref{eq:Zlambda_omega}), the renormalization factor $Z_g^2$ in the asymptotic limit ($g_0 \rightarrow 0$) of the mass-independent renormalization scheme ($m \rightarrow 0$) takes the following form:
\begin{equation}
    Z_g^2=\left[1+g_0^2\left(\nu_{R\rm}^{(1)}-\nu^{(1)}\right)+g_0^4\left(\nu_{\rm R}^{(2)}-\nu^{(2)}\right)+ g_0^4\left(\omega_{\rm R}^{(1)}-\omega^{(1)}\right) \lambda \frac{\partial \nu_{\rm R}^{(1)}}{\partial \lambda} + \mathcal{O}(g_0^6,m^2) \right]_{\substack{\lambda=\lambda_0, \\ m=m_q}}
\end{equation}
with $\nu_{\rm R}^{(\ell)} (p,m,\lambda)$, $\nu^{(\ell)} (p,m_q,\lambda_0)$, $\omega_{\rm R}^{(\ell)} (p,m)$ and $\omega^{(\ell)} (p,m_q)$ given by Eqs.~\ref{eq:nu_R_ell}), (\ref{eq:nu_L_ell}), (\ref{eq:omega_R_ell}), and (\ref{eq:omega_L_ell}) respectively.
Although $\nu_{\rm R}^{(1)}(p, m, \lambda)$ is required for a general gauge $\lambda$, in all other cases, we can choose the Feynman gauge, $\lambda = \lambda_0 = 1$. This choice significantly simplifies the computations, particularly in lattice regularization.

The amplitudes $\nu_{\rm R}^{(1)} (p,0,\lambda), \omega_{\rm R}^{(1)} (p,0), \nu_{\rm R}^{(2)} (p,0,\lambda)$ calculated in dimensional regularization for the massless case, have been already known in the literature \cite{Ellis1984,Christou:1998ws}. Moreover, the amplitudes $\nu_0^{(1)} (p,0,\lambda_0)$, $\omega_0^{(1)} (p,0)$ and $\nu_0^{(2)} (p,0,\lambda_0)$  have been calculated before using a variety of lattice actions by  \cite{Luscher1995, Skouroupathis:2007mq, Christou:1998ws, Bode:2001uz}.

Finally, the relation between the renormalized running coupling $g$ and the bare running coupling $g_0$ can be expressed as
\begin{equation} 
\begin{split}
    g^2 = \Bigg\lbrace &g^2_0 - g^4_0 \left(\nu_{\rm R}^{(1)} - \nu^{(1)}  \right) + g^6_0 \left[ \left(\nu_{\rm R}^{(1)} - \nu^{(1)} \right)^2 - \nu_{\rm R}^{(2)} + \nu^{(2)} - \lambda \frac{\partial \nu_{\rm R}^{(1)} }{\partial \lambda} \left( \omega_{\rm R}^{(1)} - \omega^{(1)} \right) \right] + \mathcal{O}(g^8_0) \Bigg\rbrace_{\substack{\lambda=\lambda_0, \\ m=m_q}}
\end{split} 
\end{equation} 
Writing $b_g(g_0)$ as an expansion in powers of $g_0$ in  Eq.~(\ref{eq:modified_g0}),
\begin{equation}
    \Tilde{g}_0^2= g_0^2 \left[1 + a m_q \left(b_g^{(1)} g_0^2+ b_g^{(2)} g_0^4 + \mathcal{O} (g_0^6)\right) \right]
    \label{eq:modified_g0_expand}
\end{equation}
we can reparametrize the bare coupling constant so that the fermion mass will be decoupled from the renormalized running coupling:
\begin{equation} 
\begin{split} 
g^2 = \Bigg\lbrace \Tilde{g}^2_0 
- \Tilde{g}^4_0 \ \bigg[ \Big(\nu_{\rm R}^{(1)}  &- \nu^{(1)}_0 \Big) + a m_q  \Big(b_g^{(1)} - \nu^{(1)}_1 \Big) \bigg] 
+ \Tilde{g}^6_0 \Bigg[ {\Big(\nu_{\rm R}^{(1)} - \nu^{(1)}_0  \Big)}^2 - \nu_{\rm R}^{(2)} + \nu^{(2)}_0  - \lambda \frac{\partial \nu_{\rm R}^{(1)} }{\partial \lambda}  \Big( \omega_{\rm R}^{(1)}  - \omega^{(1)}_0  \Big) 
\\
& + a m_q \left(  \nu^{(2)}_1 -  b_g^{(2)} + 2 \left(\nu^{(1)}_0 - \nu^{(1)}_R \right) \left(\nu^{(1)}_1 - b_g^{(1)} \right) + \ \lambda \frac{\partial \nu_{\rm R}^{(1)} }{\partial \lambda}  \omega^{(1)}_1 \right) \Bigg]+ \mathcal{O} (\Tilde{g}^8_0) \Bigg\rbrace_{\substack{\lambda=\lambda_0, \\ m=m_q}}
\end{split}
\label{eq:renorm_running_coupling}
\end{equation} 

Therefore, $b_g^{(1)}$ is given by
\begin{equation}
    b_g^{(1)} = \nu^{(1)}_1
    \label{eq:bg1_def}
\end{equation}
and at two-loop order, $b_g^{(2)}$ must be
\begin{equation}
    b_g^{(2)} 
    =  \nu^{(2)}_1 + \ \lambda \ \frac{\partial \nu_R^{(1)} }{\partial \lambda} \ \omega^{(1)}_1 \Bigg|_{\substack{\lambda=\lambda_0, \\ m=m_q}} 
    \label{eq:bg2_def}
\end{equation}

\section{One-loop Results}
\label{sec:1loop}

\begin{figure}[ht]
\centering
        \includegraphics[scale=0.70]{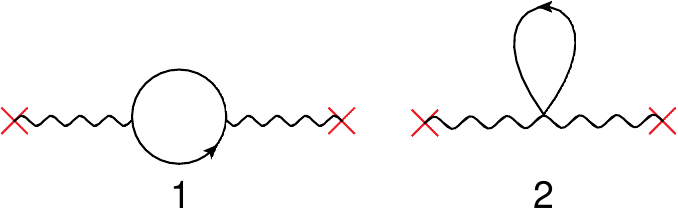}
\caption{ One-loop Feynman diagrams for fermion contributions to $\Gamma^{\rm BB, 1loop}_{\rm L, F}$. A solid line represents quarks. Wavy lines ending on a cross represent background gluons. Each diagram is meant to be symmetrized over the color indices, Lorentz indices, and momenta of the two external background fields.}
\label{fig:oneloop}
\end{figure}

As already discussed, the most efficient way of calculating $Z_g^2$ is to consider the 2-point (2-pt) Green's function of the background field. We are only interested in calculating diagrams having a fermion propagator as seen in Fig.\ref{fig:oneloop}, since these are associated with the $\mathcal{O}(am)$ effects of the strong coupling constant.

The one-loop result of the fermion contributions (``F'') to 2-pt lattice Green's function is
\begin{equation}
\Gamma^{\rm BB, 1loop}_{\rm L, F} (p,m_q)_{\mu \nu}^{a b}=\delta^{a b} \left(\delta_{\mu \nu} p^2  -p_\nu p_\mu \right) \left\{F_1(a p) + F_2\left(\frac{m_q^2}{p^2}\right) + a m_q \left[F_3 (a p) + F_4 \left(\frac{m_q^2}{p^2}\right) \right] + \mathcal{O}\left(a^2 m_q^2\right)
\right\},
\label{eq:Greens_function_oneloop_l}
\end{equation}
where
\begin{eqnarray}
F_1(a p) &=& N_f \left\{-0.0137322+0.0050467\,c_{\rm sw}-0.0298435\,c_{\rm sw}^2+\frac{2}{3}\frac{1}{16\pi^2}\ln(a^2 p^2) \right\}\nonumber\\\nonumber
F_2\left(\frac{m^2_q}{p^2}\right) &=& \frac{N_f}{16 \pi^2} \left\{\frac{8}{3} \frac{m^2_q}{p^2} - \frac{8}{3} \left(-\frac{1}{2} +\frac{m^2_q}{p^2} \right) \sqrt{1 + 4\frac{m^2_q}{p^2}} \, \archcoth\left(\sqrt{1 + 4\frac{m^2_q}{p^2}}\right) +\frac{2}{3}\ln\left(\frac{m^2_q}{p^2}\right)\right\}\nonumber\\\nonumber 
F_3(a p) &=& N_f \left\{0.0272837-0.0223503 \, c_{\rm sw} + 0.0070667 \, c_{\rm sw}^2 - (1 -  c_{\rm sw})\frac{2}{16\pi^2}\ln\left(a^2 p^2\right)\right\}\nonumber\\\nonumber
F_4 \left(\frac{m^2_q}{p^2}\right) &=& \frac{N_f}{16 \pi^2} \Bigg\{-4 \frac{m^2_q}{p^2} + 4 \left[(-1 + c_{\rm sw}) \left(1 + 4\frac{m^2_q}{p^2}\right) + 4\left(\frac{m^2_q}{p^2}\right)^2 \right] \frac{\archcoth\left(\sqrt{1 + 4\frac{m^2_q}{p^2}}\right)}{\sqrt{1 + 4\frac{m^2_q}{p^2}}} \nonumber\\\nonumber
&&\hspace{1.2cm}- (1 -  c_{\rm sw})2\ln\left(\frac{m^2_q}{p^2}\right)\Bigg\}
\label{eq:F_definitions}
\end{eqnarray}
It is worth mentioning that the errors on our lattice one-loop expressions are smaller than the last shown digit.

Since the above Green's function stems from diagrams of closed fermion loops, the one-loop results are independent of the Symanzik coefficients. Furthermore, incorporating stout links into the lattice action we find that the one-loop outcome is independent of the stout coefficient $w$. This stems from the fact that after $N$ successive smearing steps, the transverse part of a gluon field in one-loop corrections is multiplied by a factor $(1 - w a^2 \hat{p}^2)^N$ \cite{Constantinou:2022aij}. Eliminating terms of order $\mathcal{O}(a^2)$ should result in no contribution from the stout-smearing coefficient, as indicated by our findings. 

In terms of the amplitudes $\nu^{(1)}(p,m_q,\lambda_0)$ in Eq.~(\ref{eq:nu_L_ell}), using Eq.~(\ref{eq:Greens_function_oneloop_l}) we find the following:
\begin{align}
    \nu^{(1)}_0(p,0,\lambda_0)&=\nu^{(1)}_0(p,0,\lambda_0)|_{N_f=0} + F_1(a p)  \nonumber\\
    \Delta\nu^{(1)}_0(p,m_q,\lambda_0)&= \mathcal{O}\left(\frac{m_q^2}{p^2}\right) \nonumber\\
    \nu^{(1)}_1(p,0,\lambda_0)&= F_3(a p) \nonumber\\
    \Delta\nu^{(1)}_1(p,m_q,\lambda_0)&= \mathcal{O}\left(\frac{m_q^2}{p^2}\right)
    \label{eq:nu_1loop}
\end{align}

Amplitudes $\nu_0^{(1)}$ and $\nu_1^{(1)}$ can be derived using two approaches; first, by evaluating the momentum integrals and subsequently performing a first-order Taylor expansion in mass, as outlined in this section, and second, by reversing the steps. Both methods yield identical results, with the latter being the most straightforward for lattice calculations. Consequently, for the two-loop calculations, we will employ only the second approach.

Using Eq.~(\ref{eq:bg1_def}) we find the first coefficient of $b_g(g^2_0)$ as
\begin{equation}
    b_g^{(1)} =  N_f \left\{0.0272837-0.0223503 \, c_{\rm sw}+ 0.0070667 \, c_{\rm sw}^2 - (1 -  c_{\rm sw})\frac{2}{16\pi^2}\ln\left(a^2 p^2\right)\right\}+ \mathcal{O}\left(\frac{m_q^2}{p^2}\right)
\label{bg1}
\end{equation}

We stress that $b_g^{(1)}$ is independent of the gluon action used and the number of colors. 

For the tree-level Sheikholeslami-Wohlert coefficient, $c_{\rm sw}=1+\mathcal{O}(g_0^2)$, we obtain the well-established result \cite{Luscher1996}
\begin{equation}
    b_g^{(1)} = 0.012000 N_f
\end{equation}

To incorporate a mass-dependent running coupling, it is necessary to consider the terms of  $\mathcal{O}\left(m_q^2/p^2\right)$ in the amplitudes of Eq.~(\ref{eq:nu_1loop}). For heavy fermions, taking the limit $z \to \infty$ (where $z \equiv m_q/p$) and setting ${c_{\rm sw}}=1+\mathcal{O}(g_0^2)$ , we obtain
\begin{align}
    \Delta\nu^{(1)}_0(p,m_q,\lambda_0)&= \frac{N_f}{16 \pi^2} \left\{ \frac{10}{9} + \frac{4}{3} \ln(z) + \frac{2}{15z^2}   \right\}+\mathcal{O}\left(\frac{1}{z^4}\right) \nonumber\\
    \Delta\nu^{(1)}_1(p,m_q,\lambda_0)&= \frac{N_f}{16 \pi^2} \left\{ -\frac{2}{3} + \frac{2}{15z^2}  \right\}+\mathcal{O}\left(\frac{1}{z^4}\right) 
    \label{eq:nu_massive_1loop}
\end{align}
Using the above relations and Eq.~(\ref{eq:renorm_running_coupling}), one can find the logarithmic mass dependence of the renormalized running coupling in the limit of heavy fermions, as determined in Ref.~\cite{Sint1996}.

\section{Two-loop Calculations}
\label{sec:2loop}

\begin{figure}[ht!]
\centering
\includegraphics[scale=1.00]{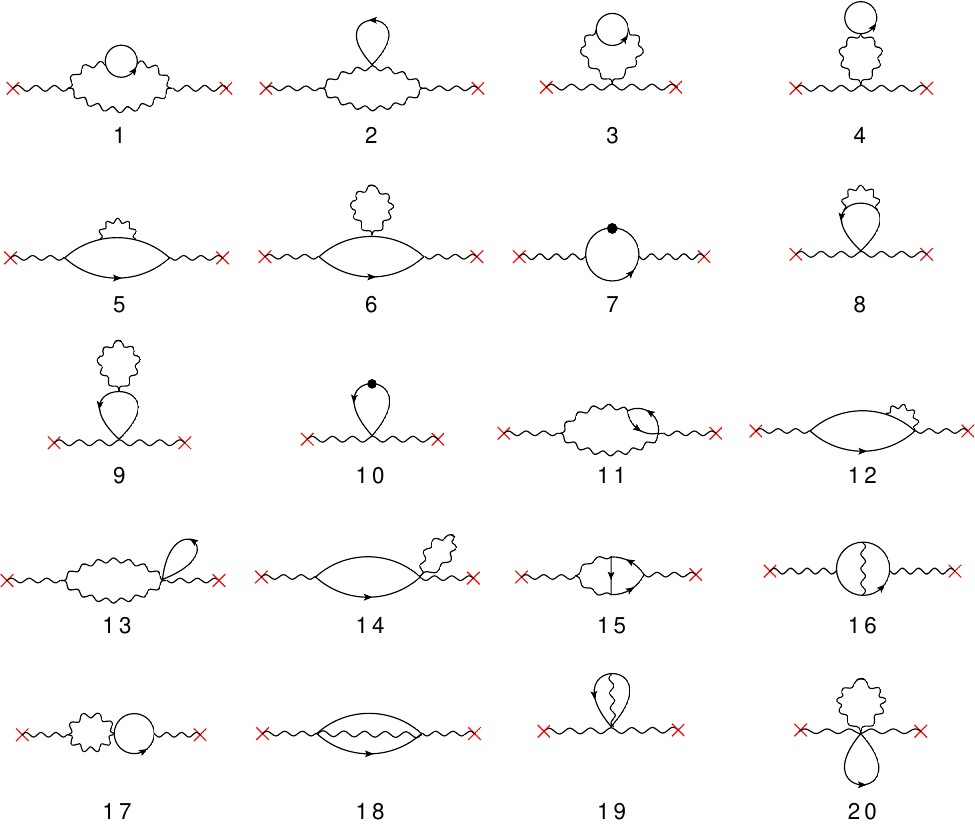}
\caption{
Two-loop Feynman diagrams for the fermion contributions to  $\Gamma^{\rm BB, 2loop}_{\rm L, F}$.
A wavy (solid) line represents gluons (quarks). Wavy lines ending on a cross represent background gluons. A solid circle is the one-loop fermion mass counterterm.
Each diagram is meant to be symmetrized over the color indices, Lorentz indices, and momenta of the two external background fields.
}
\label{fig:twoloop}
\end{figure}

To derive the two-loop amplitude $\nu^{(2)}$ of the fermion contribution, we compute the 2-point lattice Green's function of 20 Feynman diagrams, as shown in Fig.~\ref{fig:twoloop}.

The tree-level fermion propagator in momentum space is given by
\begin{equation}
\langle \psi  \bar\psi \rangle = \frac{-i \,\pcircslash + M(p,m)}{\pcirc^2 + M(p,m)^2}, 
\label{eq:propF}
\end{equation}
where $\pcircslash = \,\sum_\mu\gamma_\mu  \,\frac{1}{a}\sin(a p_\mu)$ and $M(p,m) = m + \frac{2}{a} \sum_\mu \sin^2(a p_\mu/2)$.

As mentioned in Sec.~\ref{sec:1loop}, we proceed by first expanding the denominator of Eq.~(\ref{eq:propF}) with respect to mass up to $\mathcal{O}(a)$ corrections:
\begin{equation}  
\frac{1}{\pcirc^2 + M(p,m)^2} = \frac{1}{\pcirc^2 + M(p,0)^2} \left(1 - \frac{4 m \frac{1}{a} \sum_\mu \sin^2(a p_\mu/2) }{\pcirc^2 + M(p,0)^2} + \mathcal{O}(a^2 m^2)  \right) 
\label{eq:propFm}
\end{equation}   

One main difficulty in this computation, as compared to the massless case, stems from the fact that the fermion propagator now contains contributions of $\mathcal{O}(p^{-2})$; this amplifies the presence of potential IR divergences, which must be carefully addressed. Also, the sheer number of terms which must be integrated over the two loop momenta is of the order of $\sim 10^6$; this has necessitated the creation of special-purpose integration routines, to overcome the severe constraints on CPU and memory. A brief description of the integration procedure and the associated error estimation is sketched in Appendix~\ref{ap:integration}. Computational challenges were further amplified by the distinct methodologies and manipulations required for each diagram. Particularly, the ``diamond" diagrams (diagrams 15 and 16 in Fig.~\ref{fig:twoloop}) stand out as the most challenging within this computation.

The two-loop results regarding fermion contributions to the 2-point lattice Green's function take the following form:
\begin{equation}
\sum_{\rho} \Gamma^{\rm BB, 2loop}_{\rm L, F} (p,m_q)_{\rho \rho}^{a b} = \delta^{a b}\, 3 \hat{p}^2  \sum_j \left[ G_{0,j}(a p) + a m_q \, G_{1,j}(a p) + \mathcal{O}(a^2 m_q^2) \right]
\label{eq:Greens_function_twoloop_l}
\end{equation}
The index $j$ runs over the diagrams shown in Fig. \ref{fig:twoloop}. Since two-loop diagrams can have a maximum of four fermionic vertices, $G_{i,j}(a p)$ is a polynomial in $c_{\rm sw}$ of degree up to 4.

The general form of $G_{i,j}(a p)$ is
\begin{equation}
    a^2\hat{p}^2 G_{i,j}(a p) = c_{0,i,j} + c_{1,i,j} \, a^2 \sum_\mu \frac{p^4_\mu}{p^2} + a^2 p^2 \left\{ c_{2,i,j} \left( \frac{\ln(a^2 p^2)}{(4\pi)^2} \right)^2 + c_{3,i,j} \frac{\ln(a^2 p^2)}{(4\pi)^2} + c_{4,i,j} \right\}  + O(a^4 p^4)
    \label{eq:G_coeff}
\end{equation}

where the dependence of $c_{n,i,j}$ on $N_c$, $N_f$, $c_{\rm sw}$ is given by
\begin{equation}
    c_{n,i,j} = \sum_{k=0}^{4} N_f \, c^k_{\rm sw}  \left( \frac{1}{N_c} c^{(-1,k)}_{n,i,j} + N_c \, c^{(1,k)}_{n,i,j} \right).
    \label{eq:c_coeff}
\end{equation}

It is worth mentioning that specific diagrams exhibit infrared convergence only when considered in pairs: (1, 2), (3, 4). We have evaluated these pairs accordingly, ensuring careful handling to prevent divergences in intermediate results. The results for each diagram are presented in Tables \ref{tab:coeff_table_1}-\ref{tab:coeff_table_13} in Appendix~\ref{ap:tables}.  Diagrams not appearing in these tables give vanishing contributions.

We provide here a brief overview of the validation checks performed on our calculations. Several constraints on the coefficients \( c_{n,i,j} \) were imposed to verify both the algebraic expressions and the numerical results.  

In particular, for \( c_{0,i,j} \), gauge invariance requires that  
\begin{equation}  
    \sum_{j} c_{0,i,j} = 0
\end{equation}  
where we explicitly confirmed that this condition holds.  

For the Wilson gluon action with Wilson fermions, applying a Ward-identity-like procedure—where vertices with background fields at zero momentum are expressed in terms of appropriate derivatives of inverse propagators—leads to the following additional constraints:  
\begin{align}  
\begin{split}  
    &2c_{0,i,20} + c_{0,i,14} = 0, \\  
    &c_{0,i,20} + c_{0,i,14} + c_{0,i,9} + c_{0,i,6} = 0, \\  
    &(c_{0,i,19} + c_{0,i,18})\big|_{N_c^2=2} = 0, \\ 
    &c_{0,i,4} + c_{0,i,2} + c_{0,i,3} + \frac{1}{2} c_{0,i,15} + c_{0,i,1} = 0, \\  
    &c_{0,i,11} = 0, \\  
    &c_{0,i,7} + c_{0,i,10} = 0, \\  
    &(c_{0,i,8} + c_{0,i,5})(N_c^2 - 2) + c_{0,i,12}(N_c^2 - 1) - c_{0,i,16}(N_c^2 - 1)(N_c^2 - 2) = 0.
\end{split}  
\end{align}  

These relations hold for any value of the quark mass \( m_q \) and Wilson parameter \( r \). While Ref.~\cite{Christou:1998ws} previously verified them in the massless case, we ensure their validity for the massive case as well. By substituting our numerical results for the coefficients into each identity, we consistently obtain zero within the estimated errors. 

Furthermore, as Lorentz invariance is restored in the continuum limit, the sum of terms involving the coefficients \( c_{1,i,j} \) must vanish. Among all contributions, only diagrams 1 and 2 yield nonzero values, and we have verified that their combined sum is zero.  

The coefficients \( c_{2,0,j} \) were also confirmed to align with their continuum counterparts:  
\begin{equation}
    c_{2,0,16} = \frac{1}{3} \frac{1}{N_c} N_f \,, \quad \quad
    c_{2,0,15} = \frac{4}{3} N_c N_f\,, \quad \quad
    c_{2,0,1} = -\frac{5}{3} N_c N_f\,, \quad \quad
    c_{2,0,5} = \frac{1}{3} \frac{N_c^2 - 1}{N_c} N_f
\end{equation}
For all other diagrams, the coefficients of the double logarithms vanish.  

Finally, the total contribution from single logarithms must match the continuum result:  
\begin{equation}
\sum_j c_{3,0,j} = \frac{1}{(4\pi)^2} \left( 3N_c - \frac{1}{N_c} \right) N_f
\end{equation}
This condition was verified both algebraically and numerically. 

\medskip 
To determine the second coefficient of \( b_g(g_0^2) \) as given by Eq.~(\ref{eq:bg2_def}), we use Eq.~(\ref{eq:nu_R_ell}) along with the result for the massless case of \( \nu_{\rm R}^{(1)}(p,0,\lambda) \), found, e.g., in Ref.~\cite{Christou:1998ws}. Additionally, it is important to note that the fermionic contributions to \( \omega^{(1)}(p, m_q) \)  are identical to those in \( \nu^{(1)}(p, m_q) \), i.e., \( \omega^{(1)}_1 = \nu^{(1)}_1 \). Therefore, the second coefficient \( b_g^{(2)} \) is expressed as
\begin{equation}
    b_g^{(2)} = \sum_j G_{1,j}(a p) - \frac{2 N_c}{16 \pi^2} F_3(a p) + \mathcal{O}\left( \frac{m_q^2}{p^2} \right)
    \label{eq:bg2_general_solution}
\end{equation}
where $G_{1,j}(a p)$ is defined in Eq.~(\ref{eq:G_coeff}), and it depends on the coefficients $c_{n,i,j}$ as outlined in Eq.~(\ref{eq:c_coeff}). The coefficients used in the above relation are listed in Tables~\ref{tab:coeff_table_14}–\ref{tab:coeff_table_15} in Appendix~\ref{ap:tables}. The result for the Wilson gluon action corresponds to the first line of each row in these tables and is obtained using Eq.~(\ref{eq:bg2_general_solution}):
\begin{equation}
\begin{split}
    b_g^{(2)}|_{\rm Wilson} = &\frac{N_f}{N_c}  \Bigg\{  -0.0114014(29) - 0.015674(5) c_{\rm sw} + 0.0092076(20) c_{\rm sw}^2 + 0.00035230(5) c_{\rm sw}^3 - 0.000000475(4) c_{\rm sw}^4 \\
    & \quad \quad \; + \left( 0.0767570(8) + 0.1104702(14) c_{\rm sw} - 0.0408221(18) c_{\rm sw}^2 + 0.02123107(7) c_{\rm sw}^3 \right) \frac{\ln(a^2 p^2)}{16 \pi^2} \\
    & \quad \quad \; + (1 - c_{\rm sw}) \left( \frac{\ln(a^2 p^2)}{16 \pi^2} \right)^2  \Bigg\}   \\
    & + N_f N_c \Bigg\{ -0.005463(5) + 0.003539(4) c_{\rm sw} - 0.0019859(13) c_{\rm sw}^2 - 0.00011429(12) c_{\rm sw}^3 + 0.000017174(4) c_{\rm sw}^4 \\
    & \quad \quad \quad \quad \; \, + \left( -0.156047(20) - 0.090502(20) c_{\rm sw} + 0.066449(5) c_{\rm sw}^2 - 0.0175896599(11) c_{\rm sw}^3 \right) \frac{\ln(a^2 p^2)}{16 \pi^2} \\
    & \quad \quad \quad \quad \; \, +(1 - c_{\rm sw}) \left( \frac{\ln(a^2 p^2)}{16 \pi^2} \right)^2 \Bigg\}  + \mathcal{O}\left( \frac{m_q^2}{p^2} \right)
\end{split}
\label{bg2Wilson}
\end{equation}

\medskip
\noindent Similarly, we can extract the result for the tree-level Symanzik gluon action, which is given by
\begin{equation}
\begin{split}
    b_g^{(2)}|_{\rm TLS} = &\frac{N_f}{N_c}  \Bigg\{  -0.0051236(24) - 0.013840(4) c_{\rm sw} + 0.0083484(19) c_{\rm sw}^2 + 0.00029588(7) c_{\rm sw}^3 - 0.000001975(4) c_{\rm sw}^4 \\
    & \quad \quad \; + \left( 0.0576153(9) + 0.0757940(14) c_{\rm sw} - 0.0367695(17) c_{\rm sw}^2 + 0.01954606(7) c_{\rm sw}^3 \right) \frac{\ln(a^2 p^2)}{16 \pi^2} \\
    & \quad \quad \; + (1 - c_{\rm sw}) \left( \frac{\ln(a^2 p^2)}{16 \pi^2} \right)^2  \Bigg\}   \\
    & + N_f N_c \Bigg\{ -0.007598(6) + 0.003048(6) c_{\rm sw} - 0.002202(5) c_{\rm sw}^2 - 0.00002829(13) c_{\rm sw}^3 + 0.0000156943(35) c_{\rm sw}^4 \\
    & \quad \quad \quad \quad \; \, + \left( -0.130706(33) - 0.056392(19) c_{\rm sw} + 0.059879(5) c_{\rm sw}^2 - 0.0165255904(23) c_{\rm sw}^3 \right) \frac{\ln(a^2 p^2)}{16 \pi^2} \\
    & \quad \quad \quad \quad \; \, +(1 - c_{\rm sw}) \left( \frac{\ln(a^2 p^2)}{16 \pi^2} \right)^2 \Bigg\}  + \mathcal{O}\left( \frac{m_q^2}{p^2} \right)
\end{split}
\label{bg2TLS}
\end{equation}

\medskip
\noindent while for the Iwasaki gluon action, the second coefficient $b_g^{(2)}$ becomes
\begin{equation}
\begin{split}
    b_g^{(2)}|_{\rm Iwasaki} = &\frac{N_f}{N_c}  \Bigg\{  0.0031399(23) - 0.0112507(35) c_{\rm sw} + 0.0071443(16) c_{\rm sw}^2 + 0.00021121(5) c_{\rm sw}^3 - 0.0000035152(33) c_{\rm sw}^4 \\
    & \quad \quad \; + \left( 0.0370497(14) + 0.0368072(15) c_{\rm sw} - 0.0295570(18) c_{\rm sw}^2 + 0.01625262(8) c_{\rm sw}^3 \right) \frac{\ln(a^2 p^2)}{16 \pi^2} \\
    & \quad \quad \; + (1 - c_{\rm sw}) \left( \frac{\ln(a^2 p^2)}{16 \pi^2} \right)^2  \Bigg\}   \\
    & + N_f N_c \Bigg\{ -0.012841(8) + 0.001980(9) c_{\rm sw} - 0.002891(17) c_{\rm sw}^2 + 0.00013071(11) c_{\rm sw}^3 + 0.0000129615(31) c_{\rm sw}^4 \\
    & \quad \quad \quad \quad \; \, + \left( -0.09865(8) - 0.017291(22) c_{\rm sw} + 0.048057(5) c_{\rm sw}^2 - 0.01418348(5) c_{\rm sw}^3 \right) \frac{\ln(a^2 p^2)}{16 \pi^2} \\
    & \quad \quad \quad \quad \; \, +(1 - c_{\rm sw}) \left( \frac{\ln(a^2 p^2)}{16 \pi^2} \right)^2 \Bigg\}  + \mathcal{O}\left( \frac{m_q^2}{p^2} \right)
\end{split}
\label{bg2Iwasaki}
\end{equation}

Note that in Eqs.~(\ref{bg2Wilson})--(\ref{bg2Iwasaki}) only the tree-level value of $c_{\rm sw}$ contributes. Further contributions in these equations arise from $b_g^{(1)}$ [Eq.~(\ref{bg1})] once one inserts the perturbative expansion for $c_{\rm sw}$. For completeness, we present the one-loop results for $c_{\rm sw}$: 
\begin{equation}
c_{\rm sw} \;=\; 1 + g_0^2 \, c_{\rm sw}^{(1)} ,
\end{equation}
where the one-loop coefficients~\cite{Aoki:2003sj, Horsley:2008ap, Luscher:1996vw, Aoki:1998qd} (for $N_c=3$) are given as
\begin{equation}
c_{\rm sw}^{(1)} =
\begin{cases}
0.26858825(1), & \text{Wilson}, \\[6pt] 
0.19624449(1), & \text{TLS}, \\[6pt] 
0.11300591(1), & \text{Iwasaki}.
\end{cases}
\end{equation}

Using these results for $c_{\rm sw}$, the expression for $b_g$ for $N_c =3$
[cf. Eqs.~(\ref{eq:modified_g0}), (\ref{eq:modified_g0_expand})] becomes
\begin{equation}
b_g \equiv 
\begin{cases}
 N_f \left( 0.012000(1)\, g_0^2 - 0.020067(20)\, g_0^4 \right), & \text{Wilson}, \\[6pt] 
 N_f \left( 0.012000(1)\, g_0^2 - 0.025347(30)\, g_0^4 \right), & \text{TLS}, \\[6pt] 
 N_f \left( 0.012000(1)\, g_0^2 - 0.04201(6)\, g_0^4 \right), & \text{Iwasaki},
\end{cases}
\end{equation}
which no longer depends on the logarithmic terms, and therefore is independent of the momentum $p$. Comparing with the nonperturbative results of Ref.~\cite{DallaBrida:2023fpl}, we note that there remains a pronounced difference, even when our two-loop results are taken into account.

\section{Conclusions}
\label{sec:conclusions}

In this work, we have conducted a perturbative study of discretization effects proportional to the quark mass, $\mathcal{O}(a m_q)$, on the QCD $\beta$-function.  Our analysis focuses on computing the fermionic contribution to the renormalization factor of the coupling constant, $Z_g$, using clover fermions and a family of Symanzik-improved gluons. To achieve this, we employ the background field method, which allows us to relate the renormalization of the background field propagator (2-pt Green's function) to $Z_g$. We compute all one-loop and two-loop Feynman diagrams that include at least one fermion loop.  From these contributions, we determine the coefficients $b_g^{(1)}$ and $b_g^{(2)}$, which are the one- and two-loop corrections stemming from discretization errors proportional to the quark mass. 

Our results, which treat the number of colors ($N_c$) and flavors ($N_f$) as free parameters, are particularly relevant for lattice simulations involving heavy quarks, where large values of $a m_q$ introduce non-negligible lattice artifacts. Our findings confirm that the inclusion of mass-dependent corrections in the $\beta$-function is crucial for reducing these effects and achieving controlled continuum extrapolations, particularly in precision studies of the strong coupling constant.

A natural extension of this work involves incorporating stout-smeared links into the fermion action to reduce discretization effects and statistical noise in simulations. Additionally, we aim to compute contributions from pure-gluon Feynman diagrams ($\nu^{(2)}$) using the Symanzik-improved gluon action. This will offer a more comprehensive understanding of the coupling constant’s running and play an important role in determining nonperturbative quantities in lattice QCD.

Beyond these specific extensions, further generalizations of our approach can be explored along similar lines. This includes investigating alternative gauge actions and fermion discretizations to evaluate their impact on renormalization of the gauge coupling.

\begin{acknowledgments}
 It is a pleasure to thank Mattia Dalla Brida for fruitful discussions. The project is implemented under the program of social cohesion``THALIA 2021--2027'' co-funded by the European Union through the Research and Innovation Foundation (RIF). 
\end{acknowledgments}

\section*{Data Availability}
The data that support the findings  of this article  are not publicly available upon  publication because it is not technically feasible  and/or the cost of  preparing, depositing,  and  hosting the data would be prohibitive within the terms of this research project. The  data are available  from the authors upon reasonable request.

\bibliography{references}

@article{Sheikholeslami1985,
title = {Improved continuum limit lattice action for QCD with wilson fermions},
journal = {Nucl. Phys.},
volume = {B259},
number = {4},
pages = {572-596},
year = {1985},
issn = {0550-3213},
doi = {https://doi.org/10.1016/0550-3213(85)90002-1},
url = {https://www.sciencedirect.com/science/article/pii/0550321385900021},
author = {B. Sheikholeslami and R. Wohlert},
}

@article{Horsley:2004mx,
    author = "Horsley, R. and Perlt, H. and Rakow, Paul E. L. and Schierholz, G. and Schiller, A.",
    collaboration = "QCDSF Collaboration",
    title = "{One-loop renormalisation of quark bilinears for overlap fermions with improved gauge actions}",
    eprint = "hep-lat/0404007",
    archivePrefix = "arXiv",
    reportNumber = "DESY-04-021, EDINBURGH-2004-03, LEIPZIG-LU-ITP-2004-007, LIVERPOOL-LTH-618",
    doi = "10.1016/j.nuclphysb.2005.01.044",
    journal = "Nucl. Phys.",
    volume = "B693",
    pages = "3--35",
    year = "2004",
    note = "[Nucl.Phys. B713, 601E (2005)]"
}

@article{Ellis1984,
title = {Two-loop corrections to the $\Lambda$ parameters of one-plaquette actions},
journal = {Nuclear Physics B},
volume = {235},
number = {1},
pages = {93-114},
year = {1984},
issn = {0550-3213},
doi = {https://doi.org/10.1016/0550-3213(84)90150-0},
url = {https://www.sciencedirect.com/science/article/pii/0550321384901500},
author = {R.K. Ellis and G. Martinelli}
}

@article{Luscher1995background,
    author = "Luscher, Martin and Weisz, Peter",
    title = "{Background field technique and renormalization in lattice gauge theory}",
    eprint = "hep-lat/9504006",
    archivePrefix = "arXiv",
    reportNumber = "MPI-PHT-95-27, DESY-95-056",
    doi = "10.1016/0550-3213(95)00346-T",
    journal = "Nucl. Phys.",
    volume = "B452",
    pages = "213--233",
    year = "1995"
}

@article{Luscher1995,
    author = "Luscher, Martin and Weisz, Peter",
    title = "{Computation of the relation between the bare lattice coupling and the MS coupling in SU(N) gauge theories to two loops}",
    eprint = "hep-lat/9505011",
    archivePrefix = "arXiv",
    reportNumber = "MPI-PHT-95-40, DESY-95-088",
    doi = "10.1016/0550-3213(95)00338-S",
    journal = "Nucl. Phys.",
    volume = "B452",
    pages = "234--260",
    year = "1995"
}

@book{Balian1981,
  title={Methods in Field Theory},
  author={Balian, R. and Zinn-Justin, J.},
  isbn={9789971830151},
  lccn={83940151},
  series={Les Houches Summer School Proceedings Series},
  url={https://www.worldscientific.com/doi/abs/10.1142/0004},
  year={1981},
  publisher={North-Holland Publishing Company},
  address = {Amstersdam}
}

@article{Sint1996,
title = {The running coupling from the QCD Schrödinger functional: a one-loop analysis},
journal = {Nuclear Physics B},
volume = {465},
number = {1},
pages = {71-98},
year = {1996},
issn = {0550-3213},
doi = {https://doi.org/10.1016/0550-3213(96)00020-X},
url = {https://www.sciencedirect.com/science/article/pii/055032139600020X},
author = {Stefan Sint and Rainer Sommer}
}

@article{Alles:1996cy,
    author = "Alles, B. and Feo, A. and Panagopoulos, H.",
    title = "{The Three loop Beta function in SU(N) lattice gauge theories}",
    eprint = "hep-lat/9609025",
    archivePrefix = "arXiv",
    reportNumber = "IFUP-TH-58-96, UCY-PHY-96-9",
    doi = "10.1016/S0550-3213(97)00092-8",
    journal = "Nucl. Phys.",
    volume = "B491",
    pages = "498--512",
    year = "1997"
}

@article{Skouroupathis:2007mq,
    author = "Skouroupathis, A. and Panagopoulos, H.",
    title = "{Lambda-parameter of lattice QCD with Symanzik improved gluon actions}",
    eprint = "0709.3239",
    archivePrefix = "arXiv",
    primaryClass = "hep-lat",
    doi = "10.1103/PhysRevD.76.114514",
    journal = "Phys. Rev. D",
    volume = "76",
    pages = "114514",
    year = "2007"
}

@article{Christou:1998ws,
    author = "Christou, C. and Feo, A. and Panagopoulos, H. and Vicari, E.",
    title = "{The three loop $\beta$-function of $SU(N)$ lattice gauge theories with Wilson fermions}",
    eprint = "hep-lat/9801007",
    archivePrefix = "arXiv",
    reportNumber = "IFUP-TH-65-97",
    doi = "10.1016/S0550-3213(98)00248-X",
    journal = "Nucl. Phys.",
    volume = "B525",
    pages = "387--400",
    year = "1998",
    note = "[Nucl.Phys. B608, 479(E) (2001)]"
}

@article{Bode:2001uz,
    author = "Bode, A. and Panagopoulos, H.",
    title = "{The Three loop beta function of QCD with the clover action}",
    eprint = "hep-lat/0110211",
    archivePrefix = "arXiv",
    doi = "10.1016/S0550-3213(02)00012-3",
    journal = "Nucl. Phys.",
    volume = "B625",
    pages = "198--210",
    year = "2002"
}

@article{Constantinou:2007rm,
    author = "Constantinou, M. and Panagopoulos, H.",
    title = "{QCD with overlap fermions: Running coupling and the 3-loop beta-function}",
    eprint = "0709.4368",
    archivePrefix = "arXiv",
    primaryClass = "hep-lat",
    doi = "10.1103/PhysRevD.76.114504",
    journal = "Phys. Rev. D",
    volume = "76",
    pages = "114504",
    year = "2007"
}

@article{Tarasov1980,
        title = {The gell-mann-low function of QCD in the three-loop approximation},
        journal = {Phys. Lett.},
        volume = {93B},
        number = {4},
        pages = {429-432},
        year = {1980},
        issn = {0370-2693},
        doi = {https://doi.org/10.1016/0370-2693(80)90358-5},
        url = {https://www.sciencedirect.com/science/article/pii/0370269380903585},
        author = {O.V. Tarasov and A.A. Vladimirov and A.Yu. Zharkov},
}

@article{Constantinou:2022aij,
    author = "Constantinou, Martha and Panagopoulos, Haralambos",
    title = "{Improved renormalization scheme for nonlocal operators}",
    eprint = "2207.09977",
    archivePrefix = "arXiv",
    primaryClass = "hep-lat",
    doi = "10.1103/PhysRevD.107.014503",
    journal = "Phys. Rev. D",
    volume = "107",
    number = "1",
    pages = "014503",
    year = "2023"
}

@article{Luscher1996,
    author = "Luscher, Martin and Sint, Stefan and Sommer, Rainer and Weisz, Peter",
    title = "{Chiral symmetry and O(a) improvement in lattice QCD}",
    eprint = "hep-lat/9605038",
    archivePrefix = "arXiv",
    reportNumber = "DESY-96-086, CERN-TH-96-138, MPI-PHT-96-38",
    doi = "10.1016/0550-3213(96)00378-1",
    journal = "Nucl. Phys.",
    volume = "B478",
    pages = "365--400",
    year = "1996"
}

@article{Abbott1980,
    author = "Abbott, L. F.",
    title = "{The Background Field Method Beyond One Loop}",
    reportNumber = "CERN-TH-2973",
    doi = "10.1016/0550-3213(81)90371-0",
    journal = "Nucl. Phys.",
    volume = "B185",
    pages = "189--203",
    year = "1981"
}

@article{FlavourLatticeAveragingGroupFLAG:2024oxs,
    author = "Aoki, Y. and others",
    collaboration = "Flavour Lattice Averaging Group (FLAG) Collaboration",
    title = "{FLAG Review 2024}",
    eprint = "2411.04268",
    archivePrefix = "arXiv",
    primaryClass = "hep-lat",
    reportNumber = "CERN-TH-2024-192, FERMILAB-PUB-24-0785-T",
    month = "11",
    journal = ""
}

@article{DallaBrida:2019mqg,
    author = {Dalla Brida, Mattia and H\"ollwieser, Roman and Knechtli, Francesco and Korzec, Tomasz and Ramos, Alberto and Sommer, Rainer},
    collaboration = "ALPHA Collaboration",
    title = "{Non-perturbative renormalization by decoupling}",
    eprint = "1912.06001",
    archivePrefix = "arXiv",
    primaryClass = "hep-lat",
    reportNumber = "WUB/19-05, DESY 19-224, DESY-19-224",
    doi = "10.1016/j.physletb.2020.135571",
    journal = "Phys. Lett. B",
    volume = "807",
    pages = "135571",
    year = "2020"
}

@article{DallaBrida:2023fpl,
    author = {Dalla Brida, Mattia and H\"ollwieser, Roman and Knechtli, Francesco and Korzec, Tomasz and Sint, Stefan and Sommer, Rainer},
    collaboration = "ALPHA Collaboration",
    title = "{Heavy Wilson quarks and O(a) improvement: nonperturbative results for b$_{g}$}",
    eprint = "2401.00216",
    archivePrefix = "arXiv",
    primaryClass = "hep-lat",
    doi = "10.1007/JHEP01(2024)188",
    journal = "J. High Energy Phys.",
    volume = "2024",
    number = "01",
    pages = "188",
    year = "2024"
}

@article{Dietrich:2009ns,
    author = "Dietrich, Dennis D.",
    title = "{A mass-dependent beta-function}",
    eprint = "0908.1364",
    archivePrefix = "arXiv",
    primaryClass = "hep-th",
    doi = "10.1103/PhysRevD.80.065032",
    journal = "Phys. Rev. D",
    volume = "80",
    pages = "065032",
    year = "2009"
}

@article{DallaBrida:2022eua,
    author = {Dalla Brida, Mattia and H\"ollwieser, Roman and Knechtli, Francesco and Korzec, Tomasz and Nada, Alessandro and Ramos, Alberto and Sint, Stefan and Sommer, Rainer},
    collaboration = "ALPHA Collaboration",
    title = "{Determination of $\alpha _s(m_Z)$ by the non-perturbative decoupling method}",
    eprint = "2209.14204",
    archivePrefix = "arXiv",
    primaryClass = "hep-lat",
    reportNumber = "IFIC/22-25, WUB/22-00, DESY-22-051, CERN-TH-2022-015, HU-EP-22/04",
    doi = "10.1140/epjc/s10052-022-10998-3",
    journal = "Eur. Phys. J. C",
    volume = "82",
    number = "12",
    pages = "1092",
    year = "2022"
}

@article{Aoki:2003sj,
    author = "Aoki, Sinya and Kuramashi, Yoshinobu",
    title = "{Determination of the improvement coefficient $c_{\rm sw}$ up to one loop order with the conventional perturbation theory}",
    eprint = "hep-lat/0306015",
    archivePrefix = "arXiv",
    doi = "10.1103/PhysRevD.68.094019",
    journal = "Phys. Rev. D",
    volume = "68",
    pages = "094019",
    year = "2003"
}

@article{Horsley:2008ap,
    author = "Horsley, R. and Perlt, H. and Rakow, P. E. L. and Schierholz, G. and Schiller, A.",
    title = "{Perturbative determination of $c_{\rm sw}$ for plaquette and Symanzik gauge action and stout link clover fermions}",
    eprint = "0807.0345",
    archivePrefix = "arXiv",
    primaryClass = "hep-lat",
    reportNumber = "DESY-08-034, EDINBURGH-2008-12, LEIPZIG-LU-ITP-2008-001, LIVERPOOL-LTH-792",
    doi = "10.1103/PhysRevD.78.054504",
    journal = "Phys. Rev. D",
    volume = "78",
    pages = "054504",
    year = "2008"
}

@article{Luscher:1996vw,
    author = "Luscher, M. and Weisz, P.",
    title = "{O(a) improvement of the axial current in lattice QCD to one loop order of perturbation theory}",
    eprint = "hep-lat/9606016",
    archivePrefix = "arXiv",
    reportNumber = "DESY-96-105, MPI-PHT-96-47",
    doi = "10.1016/0550-3213(96)00448-8",
    journal = "Nucl. Phys.",
    volume = "B479",
    pages = "429--458",
    year = "1996"
}

@article{Aoki:1998qd,
    author = "Aoki, Sinya and Frezzotti, Roberto and Weisz, Peter",
    title = "{Computation of the improvement coefficient $c_{\rm sw}$ to one loop with improved gluon actions}",
    eprint = "hep-lat/9808007",
    archivePrefix = "arXiv",
    reportNumber = "MPI-PHT-98-48",
    doi = "10.1016/S0550-3213(98)00742-1",
    journal = "Nucl. Phys.",
    volume = "B540",
    pages = "501--519",
    year = "1999"
}

@article{Luscher:1985wf,
    author = "Luscher, M. and Weisz, P.",
    title = "{Efficient Numerical Techniques for Perturbative Lattice Gauge Theory Computations}",
    reportNumber = "DESY-85-075",
    doi = "10.1016/0550-3213(86)90094-5",
    journal = "Nucl. Phys.",
    volume = "B266",
    pages = "309",
    year = "1986"
}

\newpage

\appendix

\section{MOMENTUM INTEGRATION}
\label{ap:integration}

Let us briefly outline some characteristics of our procedure for integration over loop momenta. Standard routines for numerical integration cannot be directly applied for a number of reasons, primarily stemming from the presence of poles in the integrands. The poles are due not only to the gluon massless propagators, but also to the fermion propagators, given that the latter are Taylor expanded with respect to the mass, leading to massless denominators; in fact, Taylor expansion exacerbates further this situation, leading to double-pole contributions from fermion propagators.

Diagrams 1 and 2 (and similarly for diagrams 3 and 4) in Fig.~\ref{fig:twoloop}, taken separately, exhibit quadruple poles, due to the presence of two gluon propagators with the same momentum, and are thus infrared divergent. However, given that each of these two pairs renormalizes the propagator of the gluon which is attached to the inner fermion loop, gauge invariance guarantees that combining the two diagrams in each pair leads to only double poles, which are thus integrable; this property, being valid for any value of the fermion mass, will remain valid even after Taylor expanding with respect to the mass. Thus, an appropriate combination of the integrands for each pair of diagrams is performed before attempting numerical integration.

Having converted all integrands in the form of expressions which contain at worst double poles, and are therefore integrable, we apply a na\"ive discretization of the Brillouin zone, in a way corresponding to a finite hypercubic lattice of size $L$\,: i.e., for all eight components of the two momentum four-vector dimensionless integration variables $p_1$ and $p_2$, integration over the interval $[0,\,2\pi)$ is converted into a summation over the values $p_{i,\mu} = 2\pi\,j/L\ (j=0,\ldots,L-1).$ Momentum values corresponding to propagator poles: $p_1 = 0$, $p_2=0$ and/or $p_1\pm p_2=0$ are excluded from the summation. This procedure will lead to the exact result when extrapolated to an infinite lattice, $L\to\infty$; see below for a description of the extrapolation method and the estimate of its associated error. Note in passing that, for finite $L$, this procedure is not equivalent to finite-lattice perturbation theory, because zero modes, and their associated non-Gaussian functional integration, are simply left out. 

Having cast the momentum integrals in the form of eightfold nested sums, a number of essential simplifications must be effected before carrying out the summations:

$\bullet$ First, all trigonometric functions forming the integrands can now be preevaluated, since their arguments only take a finite, and very limited, set of values. This feature, which is not directly applicable in adaptive algorithms, reduces execution time significantly.

$\bullet$ Second, by virtue of a number of potential symmetries, such as reflections and changing the order of nested sums, the size of the integrands is reduced and the domain of summation is shrunk to a small hypertriangular region of the original domain. (A small complication for finite lattices is that the boundaries of this region are sets of nonzero measure, and must therefore be properly accounted for.)

$\bullet$ Third, typical integrands contain sums of ${\sim}10^6$ terms (summands). Each summand is a product of trigonometric factors and each such factor may depend on some, or all, of the eight momentum components. A substantial optimization is obtained by organizing the summands in an inverse tree structure: that is, summands whose dependence on the innermost summation index is contained inside an identical factor are grouped together, so that the factor need be evaluated only once; the same procedure is applied iteratively to outer summations. In this way, innermost summations (the most CPU intensive ones) contain typically only ${\sim} 10^2$ terms, with a resultant significant reduction in CPU time.

\medskip
We have incorporated the above optimizations in a {\it{metacode}} which converts the original \textit{Mathematica} expressions to Fortran (or C) code for the evaluation of each diagram at various values of $L$.

Once results are obtained for several hypercubic lattice sizes $L^4$ (typically up to $L=32$ is sufficient), they must be extrapolated to $L\to\infty$. 
Extrapolation to infinite lattice size is of course a source of systematic error, indeed the only such source. To estimate this error, we proceed as follows: First, different
extrapolations of our numerical results, the latter generically denoted as $r_L$\,, are performed using a broad
spectrum of functional 
forms $f^k(L)$ (around 20 of them) of the type:
\begin{equation}
f^k(L) = \sum_{i,j} e^{(k)}_{i,j}\, L^{-i}\, (\ln L)^j
\label{extrapolationFit}
\end{equation}
These forms encode the expected large-$L$ behavior of lattice sums (see, e.g., \cite{Luscher:1985wf}).
A total of $N_k$ coefficients ($e^{(k)}_{i,j}$) appear in the $k^{\rm th}$ functional
form; these coefficients are determined uniquely using the results on
$N_{k}$ lattices of consecutive size $L$. 

For the $k^{\rm th}$ such extrapolation, a deviation $d_k$ is
calculated using a range of criteria for quality of fit. 
One possible criterion, as an example, is the difference:
$d_k = \bigl|f^k(L^{^*}) - r_{L^{^*}}\bigr|$, where
$L^{^*}$ is the largest lattice size which was not used
in the determination of $e^{(k)}_{i,j}$\,.
Finally, these
deviations are used to assign weights (``grades") $d_k^{-2}/(\sum_k d_k^{-2})$ to
each extrapolation; averaging over all extrapolations $e^{(k)}_{0,0}$ with these weights produces a final value for $r_\infty$,  together with the error estimate. 
We can check the reliability of our error estimates in a number of ways: In those cases where the result for a diagram is also known analytically, this result coincides with the numerical one within the
systematic error; also, extrapolations which incorporate new data from larger lattices are compatible with previous results, again within systematic error.

For the computation at hand, the above procedure for loop integration was carried out on ${\sim}500$ CPU cores of the HPC facility at the University of Cyprus, and required a total of ${\sim}10^6$ CPU hours for completion.

\section{TABLES}
\label{ap:tables}

The coefficients $c_{n,i,j}$ for each diagram are presented in the following tables. For notation, see Eqs.~\eqref{eq:G_coeff} and \eqref{eq:c_coeff}. Diagrams not included in these tables give vanishing contributions. Each row contains three values corresponding to different Symanzik-improved actions: the Wilson action (top line), the tree-level Symanzik action (middle line), and the Iwasaki action (bottom line).

\vspace*{\fill}
\begin{table}[htbp]
\centering
\begin{tabular}{c|ccc}
$j$ & $c^{(-1,0)}_{0,0,j}$ & $c^{(-1,1)}_{0,0,j}$ & $c^{(-1,2)}_{0,0,j}$ \\
\hline
$\text{}$ & $-0.0038889(7)$ & $0.00189353(20)$ & $0.0001186798(35)$  \\
$\text{5+6+7}$ & $-0.0029436(7)$ & $0.00168290(16)$ & $0.0001045679(33)$  \\
$\text{}$ & $-0.0016175(6)$ & $0.00131969(7)$ & $0 .000078574(4)$  \\
\hline
$\text{}$ & $0.00280141(14)$ & $-0.00172238(23)$ & $-0.0000424385(31)$  \\
$\text{8+9+10}$ & $0.00231537(11)$ & $-0.00153316(20)$ & $-0.0000417397(30)$  \\
$\text{}$ & $0.00156273(4)$ & $-0.00120639(14)$ & $-0.0000349575(14)$  \\
\hline
$\text{}$ & $0.00022981(15)$ & $-0.0000802(5)$ & $0$ \\
$\text{12}$ & $-0.000101289(23)$ & $-0.0000677(4)$ & $0$ \\
$\text{}$ & $-0.0005336(5)$ & $-0.00004800(35)$ & $0$ \\
\hline
$\text{}$ & $0.00316443073(21)$ & $0$ & $0$ \\
$\text{14}$ & $0.00262027066(18)$ & $0$ & $0$ \\
$\text{}$ & $0.00193541596(13)$ & $0$ & $0$ \\
\hline
$\text{}$ & $-0.0006100(8)$ & $-0.0001310(5)$ & $-0.00007620(17)$ \\
$\text{16}$ & $-0.0006319(5)$ & $-0.0001157(4)$ & $-0.00006281(18)$ \\
$\text{}$ & $-0.0006463(6)$ & $-0.00008934(31)$ & $-0.00004359(21)$ \\
\hline
$\text{}$ & $-0.00050773(6)$ & $0$ & $0$ \\
$\text{18}$ & $-0.00049253(6)$ & $0$ & $0$ \\
$\text{}$ & $-0.00043479(6)$ & $0$ & $0$ \\
\hline
$\text{}$ & $0.00039280(6)$ & $0.000040203(15)$ & $0$ \\
$\text{19}$ & $0.00054317(7)$ & $0.000033948(12)$ & $0$ \\
$\text{}$ & $0.00070176(9)$ & $0.000024069(9)$ & $0$ \\
\hline
$\text{}$ & $-0.00158227(5)$ & $0$ & $0$ \\
$\text{20}$ & $-0.00131017(4)$ & $0$ & $0$ \\
$\text{}$ & $-0.000967735(27)$ & $0$ & $0$ \\
\end{tabular}
\caption{Coefficients $c^{(-1,k)}_{0,0,j}$. }
\label{tab:coeff_table_1}
\end{table}
\vspace*{\fill}

\newpage

\vspace*{\fill}
\begin{table}[htbp]
\centering
\begin{tabular}{c|ccc}
$j$ & $c^{(-1,0)}_{0,1,j}$ & $c^{(-1,1)}_{0,1,j}$ & $c^{(-1,2)}_{0,1,j}$ \\
\hline
$\text{}$ & $0.0093260(13)$ & $-0.00291618(30)$ & $-0.000234681(10)$  \\
$\text{5+6+7}$ & $0.0071279(12)$ & $-0.0026423(5)$ & $-0.000221480(12)$  \\
$\text{}$ & $0.0040875(10)$ & $-0.0021453(4)$ & $-0.000184887(15)$  \\
\hline
$\text{}$ & $-0.0076049(7)$ & $0.00153127(4)$ & $-0.000002757(4)$  \\
$\text{8+9+10}$ & $-0.0061647(5)$ & $0.001419222(31)$ & $0.000019165(9)$  \\
$\text{}$ & $-0.0041065(5)$ & $0.00119797(7)$ & $0.000039129(8)$ \\
\hline
$\text{}$ & $-0.0000539(8)$ & $0.0006100(11)$ & $0$ \\
$\text{12}$ & $0.00044375(28)$ & $0.0005123(10)$ & $0$ \\
$\text{}$ & $0.0011289(5)$ & $0.0003603(8)$ & $0$ \\
\hline
$\text{}$ & $-0.00509976855(25)$ & $0$ & $0$ \\
$\text{14}$ & $-0.00422280501(21)$ & $0$ & $0$ \\
$\text{}$ & $-0.00311909924(16)$ & $0$ & $0$ \\
\hline
$\text{}$ & $0.0008565(24)$ & $0.0010797(30)$ & $0.0002370(4)$  \\
$\text{16}$ & $0.0009288(20)$ & $0.0009672(30)$ & $0.0002023(4)$  \\
$\text{}$ & $0.0010164(25)$ & $0.0007670(29)$ & $0.0001460(4)$ \\
\hline
$\text{}$ & $0.00057224(6)$ & $0$ & $0$ \\
$\text{18}$ & $0.00056020(6)$ & $0$ & $0$ \\
$\text{}$ & $0.000485524(31)$ & $0$ & $0$ \\
\hline
$\text{}$ & $-0.00054513(7)$ & $-0.00030519(20)$ & $0$ \\
$\text{19}$ & $-0.00078195(10)$ & $-0.00025635(16)$ & $0$ \\
$\text{}$ & $-0.00105013(9)$ & $-0.00018026(11)$ & $0$ \\
\hline
$\text{}$ & $0.00255002(30)$ & $0$ & $0$ \\
$\text{20}$ & $0.00211152(26)$ & $0$ & $0$ \\
$\text{}$ & $0.00155963(19)$ & $0$ & $0$ \\
\end{tabular}
\caption{Coefficients $c^{(-1,k)}_{0,1,j}$. }
\label{tab:coeff_table_2}
\end{table}
\vspace*{\fill}

\vspace*{\fill}
\begin{table}[htbp]
\centering
\begin{tabular}{c|ccc}
$j$ & $c^{(1,0)}_{0,0,j}$ & $c^{(1,1)}_{0,0,j}$ & $c^{(1,2)}_{0,0,j}$ \\
\hline
$\text{}$ & $-0.0020082(11)$ & $0.00022193(21)$ & $-0.0028072(12)$ \\
$\text{1+2}$ & $-0.0022026(10)$ & $0.00023620(19)$ & $-0.0030422(11)$ \\
$\text{}$ & $-0.0024047(6)$ & $0.00025562(14)$ & $-0.0033149(8)$ \\
\hline
$\text{}$ & $0.00145603(19)$ & $-0.000215063(30)$ & $0.00238751(11)$ \\
$\text{3+4}$ & $0.00146733(19)$ & $-0.000220718(28)$ & $0.00254495(29)$ \\
$\text{}$ & $0.00145912(19)$ & $-0.000220758(9)$ & $0.0025644(4)$ \\
\hline
$\text{}$ & $0 .0038889(7)$ & $-0.00189353(20)$ & $-0.0001186798(35)$  \\
$\text{5+6+7}$ & $0 .0029436(7)$ & $-0.00168290(16)$ & $-0.0001045679(33)$  \\
$\text{}$ & $0.0016175(6)$ & $-0.00131969(7)$ & $-0.000078574(4)$  \\
\hline
$\text{}$ & $-0.00280141(14)$ & $0.00172238(23)$ & $0.0000424385(31)$  \\
$\text{8+9+10}$ & $-0.00231537(11)$ & $0.00153316(20)$ & $0.0000417397(30)$  \\
$\text{}$ & $-0.00156273(4)$ & $0.00120639(14)$ & $0.0000349575(14)$  \\
\hline
$\text{}$ & $0$ & $-0.0000573(4)$ & $0.0009552(14)$ \\
$\text{11}$ & $0.000157315(13)$ & $-0.0000646(4)$ & $0.0011115(14)$ \\
$\text{}$ & $0.000330707(26)$ & $-0.0000792(4)$ & $0.0016031(14)$ \\
\hline
$\text{}$ & $-0.00011491(7)$ & $0.00029885(28)$ & $0.00026804(9)$ \\
$\text{12}$ & $0.000050644(12)$ & $0.00026594(28)$ & $0.00024220(9)$ \\
$\text{}$ & $0.00026682(23)$ & $0.00021724(26)$ & $0.00018887(6)$ \\
\hline
$\text{}$ & $-0.00316443073(21)$ & $0$ & $0$ \\
$\text{14}$ & $-0.00262027066(18)$ & $0$ & $0$ \\
$\text{}$ & $-0.00193541596(13)$ & $0$ & $0$ \\
\hline
$\text{}$ & $0.0011051(28)$ & $0.0000436(9)$ & $-0.00011562(16)$ \\
$\text{15}$ & $0.0013138(29)$ & $0.0000337(10)$ & $-0.00011661(15)$ \\
$\text{}$ & $0.0015592(34)$ & $0.0000094(12)$ & $-0.00010168(14)$ \\
\hline
$\text{}$ & $0.00045024(4)$ & $0.0000402030(29)$ & $-0.0077682(9)$ \\
$\text{18}$ & $0.00043918(4)$ & $0.0000353047(26)$ & $-0.0064232(8)$ \\
$\text{}$ & $0.00040291(4)$ & $0.0000274085(20)$ & $-0.0046228(6)$ \\
\hline
$\text{}$ & $-0.00039280(6)$ & $-0.000160812(12)$ & $0.0071560(5)$ \\
$\text{19}$ & $-0.00054317(7)$ & $-0.000135791(10)$ & $0.0057458(4)$ \\
$\text{}$ & $-0.00070176(9)$ & $-0.000096275(7)$ & $0.00372632(30)$ \\
\hline
$\text{}$ & $0.00158227(5)$ & $0$ & $0$ \\
$\text{20}$ & $0.00131017(4)$ & $0$ & $0$ \\
$\text{}$ & $0.000967735(27)$ & $0$ & $0$ \\
\end{tabular}
\caption{Coefficients $c^{(1,k)}_{0,0,j}$. }
\label{tab:coeff_table_3}
\end{table}
\vspace*{\fill}

\vspace*{\fill}
\begin{table}[htbp]
\centering
\begin{tabular}{c|ccc}
$j$ & $c^{(1,0)}_{0,1,j}$ & $c^{(1,1)}_{0,1,j}$ & $c^{(1,2)}_{0,1,j}$ \\
\hline
$\text{}$ & $0.0038168(26)$ & $-0.0017134(28)$ & $0.00060610(29)$ \\
$\text{1+2}$ & $0.0041574(25)$ & $-0.0018097(26)$ & $0.00065729(27)$ \\
$\text{}$ & $0.0045263(22)$ & $-0.0019482(22)$ & $0.00071661(19)$ \\
\hline
$\text{}$ & $-0.00297502(18)$ & $0.00165076(33)$ & $-0.00048649(8)$ \\
$\text{3+4}$ & $-0.00299543(19)$ & $0.0016997(4)$ & $-0.00052973(9)$ \\
$\text{}$ & $-0.00296564(21)$ & $0.0017142(4)$ & $-0.00054116(10)$ \\
\hline
$\text{}$ & $-0.0093260(13)$ & $0.00291618(30)$ & $0.000234681(10)$ \\
$\text{5+6+7}$ & $-0.0071279(12)$ & $0.0026423(5)$ & $0.000221480(12)$ \\
$\text{}$ & $-0.0040875(10)$ & $0.0021453(4)$ & $0.000184887(15)$ \\
\hline
$\text{}$ & $0.0076049(7)$ & $-0.00153127(4)$ & $0.000002757(4)$ \\
$\text{8+9+10}$ & $0.0061647(5)$ & $-0.001419222(31)$ & $-0.000019165(9)$ \\
$\text{}$ & $0.0041065(5)$ & $-0.00119797(7)$ & $-0.000039129(8)$ \\
\hline
$\text{}$ & $0$ & $0.00053259(6)$ & $-0.00022873(21)$ \\
$\text{11}$ & $-0.000249029(26)$ & $0.00057339(6)$ & $-0.00025542(21)$ \\
$\text{}$ & $-0.00052711(5)$ & $0.00066218(9)$ & $-0.0003586(5)$ \\
\hline
$\text{}$ & $0.0000269(4)$ & $-0.00236349(11)$ & $-0.0004656(8)$ \\
$\text{12}$ & $-0.00022187(14)$ & $-0.00209473(19)$ & $-0.0004043(7)$ \\
$\text{}$ & $-0.00056446(25)$ & $-0.00170194(28)$ & $-0.0003003(6)$ \\
\hline
$\text{}$ & $0.00509976855(25)$ & $0$ & $0$ \\
$\text{14}$ & $0.00422280501(21)$ & $0$ & $0$ \\
$\text{}$ & $0.00311909924(16)$ & $0$ & $0$ \\
\hline
$\text{}$ & $-0.001684(6)$ & $-0.0004072(32)$ & $-0.00000900(23)$ \\
$\text{15}$ & $-0.002074(6)$ & $-0.000353(4)$ & $0$ \\
$\text{}$ & $-0.002592(6)$ & $-0.000194(4)$ & $0.00000896(22)$ \\
\hline
$\text{}$ & $-0.000558677(18)$ & $-0.00030519(4)$ & $0.0022034(4)$ \\
$\text{18}$ & $-0.000546553(5)$ & $-0.000264678(34)$ & $0.00180793(29)$ \\
$\text{}$ & $-0.000504270(6)$ & $-0.000201129(25)$ & $0.00127506(20)$ \\
\hline
$\text{}$ & $0.00054513(7)$ & $0.00122076(16)$ & $-0.0018550(5)$ \\
$\text{19}$ & $0.00078195(10)$ & $0.00102539(13)$ & $-0.00147783(35)$ \\
$\text{}$ & $0.00105013(9)$ & $0.00072102(9)$ & $-0.00094452(22)$ \\
\hline
$\text{}$ & $-0.00255002(30)$ & $0$ & $0$ \\
$\text{20}$ & $-0.00211152(26)$ & $0$ & $0$ \\
$\text{}$ & $-0.00155963(19)$ & $0$ & $0$ \\
\end{tabular}
\caption{Coefficients $c^{(1,k)}_{0,1,j}$. }
\label{tab:coeff_table_4}
\end{table}
\vspace*{\fill}

\begin{table}[htbp]
\centering
\begin{tabular}{c|ccccc}
$j$ & $c^{(-1,0)}_{4,0,j}$ & $c^{(-1,1)}_{4,0,j}$ & $c^{(-1,2)}_{4,0,j}$ & $c^{(-1,3)}_{4,0,j}$ & $c^{(-1,4)}_{4,0,j}$\\
\hline
$\text{}$ & $0.00069295(5)$ & $-0.0000201015(15)$ & $0.00059633(5)$ & $0$ & $0$\\
$\text{3+4}$ & $0.000109289(11)$ & $0.00000432917(25)$ & $-0.0000204284(22)$ & $0$ & $0$\\
$\text{}$ & $-0.000372439(27)$ & $0.0000351782(10)$ & $-0.00061565(6)$ & $0$ & $0$\\
\hline
$\text{}$ & $0.0012100(4)$ & $-0.00126395(26)$ & $0.00264223(26)$ & $-0.000499707(31)$ & $-0.000043829(4)$ \\
$\text{5+6+7}$ & $0.00089239(33)$ & $-0.00108669(21)$ & $0.00208698(22)$ & $-0.000429778(30)$ & $-0.0000380007(32)$ \\
$\text{}$ & $0.00045383(23)$ & $-0.00080328(13)$ & $0.00136482(17)$ & $-0.000317099(33)$ & $-0.0000278527(24)$ \\
\hline
$\text{}$ & $-0.00011586(4)$ & $0.000186623(13)$ & $0.0000059016(14)$ & $0$ & $0$\\
$\text{12}$ & $-0.00000559(4)$ & $0.000143805(16)$ & $0.0000046789(11)$ & $0$ & $0$\\
$\text{}$ & $0.000151014(25)$ & $0.000073876(9)$ & $0.0000029442(7)$ & $0$ & $0$\\
\hline
$\text{}$ & $-0.00106378758(4)$ & $0.000826315389(28)$ & $-0.00746086682(16)$ & $0$ & $0$\\
$\text{14}$ & $-0.000880857132(34)$ & $0.000623967525(24)$ & $-0.00546527413(13)$ & $0$ & $0$\\
$\text{}$ & $-0.000650629333(25)$ & $0.000384844011(18)$ & $-0.00313752882(10)$ & $0$ & $0$\\
\hline
$\text{}$ & $0.0020710(5)$ & $0.0053702(16)$ & $-0.0016253(5)$ & $0.000166547(11)$ & $-0.0000381711(31)$\\
$\text{16}$ & $0.0011507(5)$ & $0.0046775(15)$ & $-0.0014058(4)$ & $0.000146814(9)$ & $-0.0000308278(24)$\\
$\text{}$ & $-0.0001761(4)$ & $0.0037166(12)$ & $-0.00108699(33)$ & $0.000113649(6)$ & $-0.0000202313(14)$\\
\hline
$\text{}$ & $0.000108788(6)$ & $0$ & $0$ & $0$ & $0$\\
$\text{18}$ & $0.000105420(5)$ & $0$ & $0$ & $0$ & $0$\\
$\text{}$ & $0.000095401(7)$ & $0$ & $0$ & $0$ & $0$\\
\hline
$\text{}$ & $0$ & $0.0000201015(12)$ & $-0.00119266(7)$ & $0$ & $0$\\
$\text{19}$ & $0$ & $0.0000169738(10)$ & $-0.00095763(6)$ & $0$ & $0$\\
$\text{}$ & $0$ & $0.0000120343(7)$ & $-0.00062105(4)$ & $0$ & $0$\\
\end{tabular}
\caption{Coefficients $c^{(-1,k)}_{4,0,j}$. }
\label{tab:coeff_table_5}
\end{table}
\begin{table}[htbp]
\centering
\begin{tabular}{c|ccccc}
$j$ & $c^{(-1,0)}_{4,1,j}$ & $c^{(-1,1)}_{4,1,j}$ & $c^{(-1,2)}_{4,1,j}$ & $c^{(-1,3)}_{4,1,j}$ & $c^{(-1,4)}_{4,1,j}$\\
\hline
$\text{}$ & $-0.00113873(15)$ & $0.000152595(20)$ & $-0.00015458(4)$ & $0$ & $0$\\
$\text{3+4}$ & $-0.000153849(19)$ & $-0.0000298762(19)$ & $0.00000108952(17)$ & $0$ & $0$\\
$\text{}$ & $0.00068827(8)$ & $-0.000256895(16)$ & $0.000143724(28)$ & $0$ & $0$\\
\hline
$\text{}$ & $-0.0036156(13)$ & $0.0047311(6)$ & $-0.00148909(21)$ & $0.000280350(33)$ & $-0.000017174(4)$ \\
$\text{5+6+7}$ & $-0.0027096(11)$ & $0.00392013(18)$ & $-0.00127503(20)$ & $0.00023125(5)$ & $-0.0000156943(35)$ \\
$\text{}$ & $-0.0014840(8)$ & $0.00273999(31)$ & $-0.00099593(25)$ & $0.00016039(4)$ & $-0.0000129615(31)$ \\
\hline
$\text{}$ & $0.00003208(5)$ & $-0.00037763(16)$ & $0.00008612(13)$ & $0$ & $0$\\
$\text{12}$ & $-0.00016569(6)$ & $-0.00022383(4)$ & $0.00007138(12)$ & $0$ & $0$\\
$\text{}$ & $-0.00045518(6)$ & $0.000004225(31)$ & $0.00004785(11)$ & $0$ & $0$\\
\hline
$\text{}$ & $0.00211357914(13)$ & $-0.00365949362(15)$ & $0.00176668007(6)$ & $0$ & $0$\\
$\text{14}$ & $0.00175012503(11)$ & $-0.00276335794(12)$ & $0.00129413795(5)$ & $0$ & $0$\\
$\text{}$ & $0.00129269848(8)$ & $-0.00170435438(9)$ & $0.00074294446(4)$ & $0$ & $0$\\
\hline
$\text{}$ & $-0.0086414(26)$ & $-0.016368(5)$ & $0.0086893(20)$ & \
$0.00007195(4)$ & $0.0000166996(23)$\\
$\text{16}$ & $-0.0036992(21)$ & $-0.014615(4)$ & $0.0080105(18)$ & \
$0.00006463(4)$ & $0.0000137191(26)$\\
$\text{}$ & $0.0032268(21)$ & $-0.0119436(35)$ & $0.0070482(16)$ & \
$0.000050824(31)$ & $0.0000094464(11)$\\
\hline
$\text{}$ & $-0.000151344(20)$ & $0$ & $0$ & $0$ & $0$\\
$\text{18}$ & $-0.000145358(19)$ & $0$ & $0$ & $0$ & $0$\\
$\text{}$ & $-0.000128649(17)$ & $0$ & $0$ & $0$ & $0$\\
\hline
$\text{}$ & $0$ & $-0.000152595(16)$ & $0.00030916(6)$ & $0$ & $0$\\
$\text{19}$ & $0$ & $-0.000128174(13)$ & $0.00024631(5)$ & $0$ & $0$\\
$\text{}$ & $0$ & $-0.000090128(8)$ & $0.000157421(29)$ & $0$ & $0$\\
\end{tabular}
\caption{Coefficients $c^{(-1,k)}_{4,1,j}$. }
\label{tab:coeff_table_6}
\end{table}

\begin{table}[htbp]
\centering
\begin{tabular}{c|ccccc}
$j$ & $c^{(1,0)}_{4,0,j}$ & $c^{(1,1)}_{4,0,j}$ & $c^{(1,2)}_{4,0,j}$ & $c^{(1,3)}_{4,0,j}$ & $c^{(1,4)}_{4,0,j}$\\
\hline
$\text{}$ & $-0.0010903(6)$ & $0.00007742(9)$ & $-0.0014856(5)$ & $0$ & $0$\\
$\text{1+2}$ & $-0.0009526(6)$ & $0.00006439(9)$ & $-0.0013209(5)$ & $0$ & $0$\\
$\text{}$ & $-0.0007641(5)$ & $0.00004030(7)$ & $-0.00105896(30)$ & $0$ & $0$\\
\hline
$\text{}$ & $-0.00074284(11)$ & $0.0000377508(29)$ & $-0.00074562(7)$ & $0$ & $0$\\
$\text{3+4}$ & $-0.000130647(20)$ & $-0.0000017160(7)$ & $-0.000056858(11)$ & $0$ & $0$\\
$\text{}$ & $0.00049279(7)$ & $-0.000066534(9)$ & $0.00078204(11)$ & $0$ & $0$\\
\hline
$\text{}$ & $-0.0012100(4)$ & $0.00126395(26)$ & $-0.00264223(26)$ & $0.000499707(31)$ & $0.000043829(4)$ \\
$\text{5+6+7}$ & $-0.00089239(33)$ & $0.00108669(21)$ & $-0.00208698(22)$ & $0.000429778(30)$ & $0.0000380007(32)$ \\
$\text{}$ & $-0.00045383(23)$ & $0.00080328(13)$ & $-0.00136482(17)$ & $0.000317099(33)$ & $0.0000278527(24)$ \\
\hline
$\text{}$ & $0.0000690696(32)$ & $0.000040123(19)$ & $-0.00055809(9)$ & $0$ & $0$\\
$\text{11}$ & $0.0000694665(25)$ & $0.000041335(12)$ & $-0.00052884(9)$ & $0$ & $0$\\
$\text{}$ & $0.000069021(7)$ & $0.000043270(26)$ & $-0.00052731(19)$ & $0$ & $0$\\
\hline
$\text{}$ & $0.000057929(19)$ & $-0.00022232(10)$ & $-0.0001601650(29)$ & $-0.0000140519(33)$ & $0$\\
$\text{12}$ & $0.000002797(22)$ & $-0.00015567(8)$ & $-0.000114738(7)$ & $-0.0000146039(33)$ & $0$\\
$\text{}$ & $-0.000075507(12)$ & $-0.00007059(5)$ & $-0.000062081(25)$ & $-0.0000135287(29)$ & $0$\\
\hline
$\text{}$ & $0.00106378758(4)$ & $-0.000939614357(27)$ & $0.00880084109(16)$ & $0$ & $0$\\
$\text{14}$ & $0.000880857132(34)$ & $-0.000753148083(23)$ & $0.00699307822(13)$ & $0$ & $0$\\
$\text{}$ & $0.000650629333(25)$ & $-0.000524268946(17)$ & $0.00478649205(10)$ & $0$ & $0$\\
\hline
$\text{}$ & $0.0029688(18)$ & $0.0006307(9)$ & $-0.00027922(31)$ & $-0.000149584(28)$ & $0$\\
$\text{15}$ & $0.0040253(19)$ & $0.0007783(10)$ & $0.00004701(28)$ & $-0.00014563(5)$ & $0$\\
$\text{}$ & $0.0068251(27)$ & $0.0011528(9)$ & $0.0011132(4)$ & $-0.00014517(10)$ & $0$\\
\hline
$\text{}$ & $0$ & $-0.00016956(8)$ & $0.00200515(32)$ & $0$ & $0$\\
$\text{17}$ & $0$ & $-0.00014250(8)$ & $0.00168484(34)$ & $0$ & $0$\\
$\text{}$ & $0$ & $-0.00010318(8)$ & $0.0012198(4)$ & $0$ & $0$\\
\hline
$\text{}$ & $-0.000088930(22)$ & $0.0000004360(6)$ & $-0.000170561(22)$ & $0$ & $0$\\
$\text{18}$ & $-0.000087444(22)$ & $0.0000005690(6)$ & $-0.000111322(10)$ & $0$ & $0$\\
$\text{}$ & $-0.000082212(21)$ & $0.0000010483(6)$ & $-0.000102964(11)$ & $0$ & $0$\\
\hline
$\text{}$ & $0$ & $-0.0000050254(4)$ & $0.000298165(23)$ & $0$ & $0$\\
$\text{19}$ & $0$ & $-0.00000424346(31)$ & $0.000239407(19)$ & $0$ & $0$\\
$\text{}$ & $0$ & $-0.00000300858(21)$ & $0.000155263(12)$ & $0$ & $0$\\
\end{tabular}
\caption{Coefficients $c^{(1,k)}_{4,0,j}$. }
\label{tab:coeff_table_7}
\end{table}

\begin{table}[htbp]
\centering
\begin{tabular}{c|ccccc}
$j$ & $c^{(1,0)}_{4,1,j}$ & $c^{(1,1)}_{4,1,j}$ & $c^{(1,2)}_{4,1,j}$ & $c^{(1,3)}_{4,1,j}$ & $c^{(1,4)}_{4,1,j}$\\
\hline
$\text{}$ & $0.0022779(19)$ & $-0.0011826(14)$ & $0.00025340(12)$ & $0$ & $0$\\
$\text{1+2}$ & $0.0020174(21)$ & $-0.0010873(15)$ & $0.00021920(13)$ & $0$ & $0$\\
$\text{}$ & $0.0016425(23)$ & $-0.0009012(16)$ & $0.00016564(7)$ & $0$ & $0$\\
\hline
$\text{}$ & $0.00127531(27)$ & $-0.00028828(6)$ & $0.00018090(4)$ & $0$ & $0$\\
$\text{3+4}$ & $0.00019508(8)$ & $0.0000162243(30)$ & $0.0000173368(18)$ & $0$ & $0$\\
$\text{}$ & $-0.00098527(9)$ & $0.000537723(21)$ & $-0.000171132(33)$ & $0$ & $0$\\
\hline
$\text{}$ & $0.0036156(13)$ & $-0.0047311(6)$ & $0.00148909(21)$ & $-0.000280350(33)$ & $0.000017174(4)$ \\
$\text{5+6+7}$ & $0.0027096(11)$ & $-0.00392013(18)$ & $0.00127503(20)$ & $-0.00023125(5)$ & $0.0000156943(35)$ \\
$\text{}$ & $0.0014840(8)$ & $-0.00273999(31)$ & $0 .00099593(25)$ & $-0.00016039(4)$ & $0.0000129615(31)$ \\
\hline
$\text{}$ & $-0.000105343(12)$ & $-0.0002971(4)$ & $0.00015842(5)$ & $0$ & $0$\\
$\text{11}$ & $-0.000107988(4)$ & $-0.0003025(4)$ & $0.00014435(6)$ & $0$ & $0$\\
$\text{}$ & $-0.000111859(15)$ & $-0.00031521(34)$ & $0.00013708(8)$ & $0$ & $0$\\
\hline
$\text{}$ & $-0.000016042(26)$ & $0.00083334(26)$ & $-0.00000058(19)$ & $-0.00000181(12)$ & $0$\\
$\text{12}$ & $0.000082843(29)$ & $0.00064044(24)$ & $-0.00002039(18)$ & $0.00000152(11)$ & $0$\\
$\text{}$ & $0.000227590(28)$ & $0.00038125(23)$ & $-0.00005439(18)$ & $0.00000589(9)$ & $0$\\
\hline
$\text{}$ & $-0.00211357914(13)$ & $0.00416125932(15)$ & $-0.00208397640(6)$ & $0$ & $0$\\
$\text{14}$ & $-0.00175012503(11)$ & $0.00333545827(12)$ & $-0.00165591105(5)$ & $0$ & $0$\\
$\text{}$ & $-0.00129269848(8)$ & $0.00232182387(9)$ & $-0.00113340718(4)$ & $0$ & $0$\\
\hline
$\text{}$ & $-0.010179(5)$ & $0.003975(4)$ & $-0.0013517(12)$ & $0.000167869(18)$ & $0$\\
$\text{15}$ & $-0.010526(5)$ & $0.003424(6)$ & $-0.001634(5)$ & $0.000201442(28)$ & $0$\\
$\text{}$ & $-0.013580(8)$ & $0.001941(9)$ & $-0.002431(17)$ & $0.00028520(5)$ & $0$\\
\hline
$\text{}$ & $0$ & $0.0007509(4)$ & $-0.00047487(15)$ & $0$ & $0$\\
$\text{17}$ & $0$ & $0.0006310(4)$ & $-0.00039904(15)$ & $0$ & $0$\\
$\text{}$ & $0$ & $0.0004569(5)$ & $-0.00028895(16)$ & $0$ & $0$\\
\hline
$\text{}$ & $0.000128347(32)$ & $-0.0000039282(15)$ & $0.000010186(8)$ & $0$ & $0$\\
$\text{18}$ & $0.000126678(32)$ & $-0.0000045205(8)$ & $0.0000025432(30)$ & $0$ & $0$\\
$\text{}$ & $0.000120257(31)$ & $-0.0000076210(18)$ & $0.000017301(8)$ & $0$ & $0$\\
\hline
$\text{}$ & $0$ & $0.000038149(5)$ & $-0.000077290(19)$ & $0$ & $0$\\
$\text{19}$ & $0$ & $0.000032044(4)$ & $-0.000061576(15)$ & $0$ & $0$\\
$\text{}$ & $0$ & $0.0000225319(27)$ & $-0.000039355(9)$ & $0$ & $0$\\
\end{tabular}
\caption{Coefficients $c^{(1,k)}_{4,1,j}$. }
\label{tab:coeff_table_8}
\end{table}

\begin{table}[htbp]
\centering
\begin{tabular}{c|ccc}
$j$ & $c^{(-1,0)}_{3,0,j}$ & $c^{(-1,1)}_{3,0,j}$ & $c^{(-1,2)}_{3,0,j}$ \\
\hline
$\text{}$ & $-0.04370491067(1)$ & $0.00949408(1)$ & $0.00589886(1)$  \\
$\text{5+6+7}$ & $-0.02841748010(1)$ & $0.00850855(1)$ & $0.00524423(1)$ \\
$\text{}$ & $-0.00770276157(1)$ & $0.00675901(1)$ & $0.0041086(1)$ \\
\hline
$\text{}$ & $0.00528244554(15)$ & $-0.01449434464(6)$ & $0$ \\
$\text{12}$ & $-0.00015727738(17)$ & $-0.01261049781(6)$ & $0$ \\
$\text{}$ & $-0.00813690678(30)$ & $-0.00951504612(12)$ & $0$ \\
\hline
$\text{}$ & $0.051644463(1)$ & $0$ & $0$ \\
$\text{14}$ & $0.042763607(1)$ & $0$ & $0$ \\
$\text{}$ & $0.031586572(1)$ & $0$ & $0$ \\
\hline
$\text{}$ & $-0.019554572(13)$ & $0.005000258(35)$ & $-0.00589887(4)$ \\
$\text{16}$ & $-0.020521423(13)$ & $0.004101944(27)$ &
$-0.00524423(4)$ \\
$\text{}$ & $-0.022079477(25)$ & $0.002756031(20)$ & $-0.00410860(4)$ \\
\end{tabular}
\caption{Coefficients $c^{(-1,k)}_{3,0,j}$, \ $r$=1. }
\label{tab:coeff_table_9}
\end{table}

\begin{table}[htbp]
\centering
\begin{tabular}{c|cccc}
$j$ & $c^{(-1,0)}_{3,1,j}$ & $c^{(-1,1)}_{3,1,j}$ & $c^{(-1,2)}_{3,1,j}$ & $c^{(-1,3)}_{3,1,j}$ \\
\hline
$\text{}$ & $0.21587753242(1)$ & $-0.24558874(1)$ & $-0.057594788(1)$ & $0.02643796(1)$  \\
$\text{5+6+7}$ & $0.15018204843(1)$ & $-0.18225250(1)$ & $-0.051609539(1)$ & $0.024391931(1)$  \\
$\text{}$ & $0.06337224167(1)$ & $-0.096343704(1)$ & $-0.041069233(1)$ & $0.020346379(1)$ \\
\hline
$\text{}$ & $-0.01883007542(14)$ & $0.06150748172(8)$ & $-0.02174151702(13)$ & $0$ \\
$\text{12}$ & $-0.0037152229(6)$ & $0.0464924430(6)$ & $-0.01891574666(11)$ & $0$ \\
$\text{}$ & $0.02001163378(28)$ & $0.02318904363(19)$ & $-0.014272569341(18)$ & $0$ \\
\hline
$\text{}$ & $-0.15493390(1)$ & $0.32746670(1)$ & $0$ & $0$ \\
$\text{14}$ & $-0.12829082(1)$ & $0.24727675(1)$ & $0$ & $0$ \\
$\text{}$ & $-0.094759716(1)$ & $0.15251271(1)$ & $0$ & $0$ \\
\hline
$\text{}$ & $0.0346429(8)$ & $-0.0329152(14)$ & $0.0385142(18)$ & $-0.00520689(7)$ \\
$\text{16}$ & $0.0394393(9)$ & $-0.0357226(14)$ & $0.0337558(17)$ & $-0.00484587(7)$ \\
$\text{}$ & $0.0484256(14)$ & $-0.0425509(15)$ & $0.0257848(18)$ & $-0.00409376(8)$  \\
\end{tabular}
\caption{Coefficients $c^{(-1,k)}_{3,1,j}$. }
\label{tab:coeff_table_10}
\end{table}

\begin{table}[htbp]
\centering
\begin{tabular}{c|ccc}
$j$ & $c^{(1,0)}_{3,0,j}$ & $c^{(1,1)}_{3,0,j}$ & $c^{(1,2)}_{3,0,j}$ \\
\hline
$\text{}$ & $0.0813261088(8)$ & $-0.0252335742(7)$ & $0.14921733618(35)$ \\
$\text{1+2}$ & $0.0813261088(8)$ & $-0.0252335742(7)$ & $0.14921733618(35)$ \\
$\text{}$ & $0.0813261088(8)$ & $-0.0252335742(7)$ & $0.14921733618(35)$ \\
\hline
$\text{}$ & $0.04370491067(1)$ & $-0.00949408(1)$ & $-0.00589886(1)$ \\
$\text{5+6+7}$ & $0.02841748010(1)$ & $-0.00850855(1)$ & $-0.00524423(1)$ \\
$\text{}$ & $0.00770276157(1)$ & $-0.00675901(1)$ & $-0.0041086(1)$  \\
\hline
$\text{}$ & $0$ & $0.002523357011(34)$ & $-0.02984346720(4)$ \\
$\text{11}$ & $0$ & $0.002523357011(34)$ & $-0.02984346720(4)$ \\
$\text{}$ & $0$ & $0.002523357011(34)$ & $-0.02984346720(4)$ \\
\hline
$\text{}$ & $-0.00264122277(7)$ & $-0.005000260418(22)$ & $0.010528965436(18)$ \\
$\text{12}$ & $0.00007863869(8)$ & $-0.005199568630(25)$ & $0.00975201723(8)$ \\
$\text{}$ & $0.00406845339(15)$ & $-0.00491505806(32)$ & $0.008155014073(31)$ \\
\hline
$\text{}$ & $-0.0516445(1)$ & $0$ & $0$ \\
$\text{14}$ & $-0.0427636(1)$ & $0$ & $0$ \\
$\text{}$ & $-0.0315866(1)$ & $0$ & $0$ \\
\hline
$\text{}$ & $-0.051745(6)$ & $0.02711097(29)$ & $-0.004630101(5)$ \\
$\text{15}$ & $-0.048058(7)$ & $0.0263247(4)$ & $-0.004507790(11)$ \\
$\text{}$ & $-0.042509(8)$ & $0.0242905(8)$ & $-0.00404642(30)$ \\
\hline

$\text{}$ & $0$ & $0 .0100951(27)$ & $-0.119379(10)$ \\
$\text{17}$ & $0$ & $0 .0100951(27)$ & $-0.119379(10)$ \\
$\text{}$ & $0$ & $0 .0100951(27)$ & $-0.119379(10)$ \end{tabular}
\caption{Coefficients $c^{(1,k)}_{3,0,j}$. }
\label{tab:coeff_table_11}
\end{table}

\begin{table}[htbp]
\centering
\begin{tabular}{c|cccc}
$j$ & $c^{(1,0)}_{3,1,j}$ & $c^{(1,1)}_{3,1,j}$ & $c^{(1,2)}_{3,1,j}$ & $c^{(1,3)}_{3,1,j}$ \\
\hline
$\text{}$ & $-0.174414004(5)$ & $0.1497470687(9)$ & $-0.03533360140(4)$ & $0$ \\
$\text{1+2}$ & $-0.174414004(5)$ & $0.1497470687(9)$ & $-0.03533360140(4)$ & $0$ \\
$\text{}$ & $-0.174414004(5)$ & $0.1497470687(9)$ & $-0.03533360140(4)$ & $0$ \\
\hline
$\text{}$ & $-0.21587753242(1)$ & $0.245589(1)$ & $0.0575948(1)$ & $-0.026438(1)$ \\
$\text{5+6+7}$ & $-0.15018204843(1)$ & $0.182253(1)$ & $0.0516095(1)$ & $-0.0243919(1)$ \\
$\text{}$ & $-0.06337224167(1)$ & $0.0963437(1)$ & $0.0410692(1)$ & $-0.0203464(1)$ \\
\hline
$\text{}$ & $0$ & $-0.00167630123(4)$ & $0.007066720288(32)$ & $0$ \\
$\text{11}$ & $0$ & $-0.00167630123(4)$ & $0.007066720288(32)$ & $0$ \\
$\text{}$ & $0$ & $-0.00167630123(4)$ & $0.007066720288(32)$ & $0$ \\
\hline
$\text{}$ & $0 .00941503771(7)$ & $-0.00741987168(7)$ & $-0.043181620961(25)$ & $0 .015793448154(27)$ \\
$\text{12}$ & $0 .00185761146(29)$ & $0 .0014624225(4)$ & $-0.0408514560(5)$ & $0 .01462802585(8)$ \\
$\text{}$ & $-0.01000581689(14)$ & $0 .01435074267(33)$ & $-0.03504827712(25)$ & $0 .01223252112(9)$ \\
\hline
$\text{}$ & $0.154933(1)$ & $-0.372367(1)$ & $0$ & $0$ \\
$\text{14}$ & $0.128291(1)$ & $-0.298471(1)$ & $0$ & $0$ \\
$\text{}$ & $0.0947597(1)$ & $-0.207766(1)$ & $0$ & $0$ \\
\hline
$\text{}$ & $0.044566(20)$ & $0.0328527(13)$ & $0.05203333(23)$ & $-0.0069451508(11)$ \\
$\text{15}$ & $0.038412(33)$ & $0.0367891(11)$ & $0.04911856(31)$ & $-0.0067616854(23)$ \\
$\text{}$ & $0.02905(8)$ & $0.0426235(30)$ & $0.0420334(6)$ & $-0.00606963(5)$ \\
\hline
$\text{}$ & $0$ & $-0.111897(20)$ & $0.028270(5)$ & $0$ \\
$\text{17}$ & $0$ & $-0.101166(19)$ & $0.028270(5)$ & $0$ \\
$\text{}$ & $0$ & $-0.085583(22)$ & $0.028270(5)$ & $0$ \\
\end{tabular}
\caption{Coefficients $c^{(1,k)}_{3,1,j}$. }
\label{tab:coeff_table_12}
\end{table}

\begin{table}[htbp]
\centering
\begin{tabular}{c|cccccc}
$j$ & $c^{(-1,0)}_{2,0,j}$  & $c^{(-1,0)}_{2,1,j}$ & $c^{(-1,1)}_{2,1,j}$  &$c^{(1,0)}_{2,0,j}$  & $c^{(1,0)}_{2,1,j}$ &$c^{(1,1)}_{2,1,j}$ 
\\
\hline
$\text{}$ & $0$ &$0$ &$0$ &$-\frac{5}{3}$  & $5$ &$-5$ 
\\
$\text{1+2}$ & $0$ &$0$ &$0$ &$-\frac{5}{3}$  & $5$ &$-5$ 
\\
$\text{}$ & $0$ &$0$ &$0$ &$-\frac{5}{3}$  & $5$ &$-5$ 
\\
\hline
$\text{}$ & $-\frac{1}{3}$ &$1$ &$0$ &$\frac{1}{3}$  & $-1$ &$0$ 
\\
$\text{5+6+7}$ 
& $-\frac{1}{3}$  &$1$ &$0$ &$\frac{1}{3}$  & $-1$ &$0$ 
\\
$\text{}$ & $-\frac{1}{3}$  &$1$ &$0$ &$\frac{1}{3}$  & $-1$ &$0$ 
\\
\hline
 $\text{}$ & $0$ & $0$ & $0$ & $0$ & $0$ &$\frac{1}{2}$ 
\\
 $\text{11}$ & $0$ & $0$ & $0$ & $0$ & $0$ &$\frac{1}{2}$ 
\\
 $\text{}$ & $0$ & $0$ & $0$ & $0$ & $0$ &$\frac{1}{2}$ 
\\
\hline
 $\text{}$ & $0$ & $0$ & $0$ & $0$ & $0$ &$1$ 
\\
 $\text{12}$ & $0$ & $0$ & $0$ & $0$ & $0$ &$1$ 
\\
 $\text{}$ & $0$ & $0$ & $0$ & $0$ & $0$ &$1$ 
\\
\hline
 $\text{}$ & $0$ & $0$ & $0$ &$\frac{4}{3}$  & $-3$ &$-\frac{3}{2}$ 
\\
 $\text{15}$ & $0$ & $0$ &  $0$ &$\frac{4}{3}$  & $-3$ &$-\frac{3}{2}$ 
\\
 $\text{}$ & $0$ & $0$ &  $0$ &$\frac{4}{3}$  & $-3$ &$-\frac{3}{2}$ 
\\
\hline
 $\text{}$ & $\frac{1}{3}$  & $0$ &  $-1$ &$0$  & $0$ &$0$ 
\\
 $\text{16}$ & $\frac{1}{3}$  & $0$ & $-1$ & $0$  & $0$ &$0$ 
\\
$\text{}$ & $\frac{1}{3}$  & $0$ &  $-1$ & $0$  & $0$ &$0$ 
\\
\hline
 $\text{}$ & $0$  & $0$ &  $0$ &$0$  & $0$ &$4$ 
\\
 $\text{17}$ & $0$  & $0$ & $0$ & $0$  & $0$ &$4$ 
\\
$\text{}$ & $0$  & $0$ &  $0$ & $0$  & $0$ &$4$ \\
\end{tabular}
\caption{Coefficients $c^{(l,k)}_{2,i,j}$, $r$=1. }
\label{tab:coeff_table_13}
\end{table}

\begin{table}[htbp]
\centering
\begin{tabular}{c|ccccc}
\text{n} & $\sum_j c^{(-1,0)}_{n,1,j}$ & $\sum_j c^{(-1,1)}_{n,1,j}$ & $\sum_j c^{(-1,2)}_{n,1,j}$ & $\sum_j c^{(-1,3)}_{n,1,j}$ & $\sum_j c^{(-1,4)}_{n,1,j}$\\
\hline
$\text{}$ & $1$ & $-1$ & $0$ & $0$ & $0$ \\
$\text{2}$ & $1$ & $-1$ & $0$ & $0$ & $0$ \\
$\text{}$ & $1$ & $-1$ & $0$ & $0$ & $0$ \\
\hline
$\text{}$ & $0 .0767570(8)$ & $0 .1104702(14)$ & $-0.0408221(18)$ & \
$0 .02123107(7)$ & $0$ \\
$\text{3}$ & $0 .0576153(9)$ & $0 .0757940(14)$ & $-0.0367695(17)$ & \
$0 .01954606(7)$ & $0$ \\
$\text{}$ & $0 .0370497(14)$ & $0 .0368072(15)$ & $-0.0295570(18)$ & \
$0 .01625262(8)$ & $0$ \\
\hline
$\text{}$ & $-0.0114014(29)$ & $-0.015674(5)$ & $0.0092076(20)$ & $0.00035230(5)$ & $-0.000000475(4)$ \\
$\text{4}$ & $-0.0051236(24)$ & $-0.013840(4)$ & $0.0083484(19)$ & $0.00029588(7)$ & $-0.000001975(4)$ \\
$\text{}$ & $0 .0031399(23)$ & $-0.0112507(35)$ & $0.0071443(16)$ & $0.00021121(5)$ & $-0.0000035152(33)$ \\
\hline
\end{tabular}
\caption{Coefficients $\sum_j c^{(-1,k)}_{n,1,j}$.}
\label{tab:coeff_table_14}
\end{table}

\begin{table}[htbp]
\centering
\begin{tabular}{c|ccccc}
\text{n} & $\sum_j c^{(1,0)}_{n,1,j}$ & $\sum_j c^{(1,1)}_{n,1,j}$ & $\sum_j c^{(1,2)}_{n,1,j}$ & $\sum_j c^{(1,3)}_{n,1,j}$ & $\sum_j c^{(1,4)}_{n,1,j}$\\
\hline
$\text{}$ & $1$ & $-1$ & $0$ & $0$ & $0$ \\
$\text{2}$ & $1$ & $-1$ & $0$ & $0$ & $0$ \\
$\text{}$ & $1$ & $-1$ & $0$ & $0$ & $0$ \\
\hline
$\text{}$ & $-0.181377(20)$ & $-0.065172(20)$ & $0.066449(5)$ & $-0.0175896599(11)$ & $0$ \\
$\text{3}$ & $-0.156036(33)$ & $-0.031062(19)$ & $0.059879(5)$ & $-0.0165255904(23)$ & $0$ \\
$\text{}$ & $-0.12398(8)$ & $0 .008039(22)$ & $0.048057(5)$ & $-0.01418348(5)$ & $0$ \\
\hline
$\text{}$ & $-0.005117(5)$ & $0 .003255(4)$ & $-0.0018964(13)$ & $-0.00011429(12)$ & $0.000017174(4)$ \\
$\text{4}$ & $-0.007252(6)$ & $0 .002765(6)$ & $-0.002112(5)$ & $-0.00002829(13)$ & $0.0000156943(35)$ \\
$\text{}$ & $-0.012495(8)$ & $0 .001697(9)$ & $-0.002802(17)$ & $0 .00013071(11)$ & $0.0000129615(31)$ \\
\hline
\end{tabular}
\caption{Coefficients $\sum_j c^{(1,k)}_{n,1,j}$.}
\label{tab:coeff_table_15}
\end{table}

\end{document}